\documentclass[epsf,usegraphicx,usenatbib]{mn2e}

\usepackage{xspace}
\usepackage{aas_macros}
\usepackage{subfigure}
\usepackage{amsmath}
\usepackage{amsfonts}
\usepackage{times}

\title[Results from the first 5 years of RATS]{Stellar variability on time-scales of minutes: results from the first 5 years of the Rapid Temporal Survey (RATS)\thanks{Based on observations made with the Isaac Newton and William Herschel Telescopes operated on the island of La Palma by the Isaac Newton Group in the Spanish Observatorio del Roque de los Muchachos of the Instituto de Astrofísica de Canarias and also observations collected at the European Organisation for Astronomical Research in the Southern Hemisphere, Chile, proposal 075.D-0111.}}

\author[Barclay et al.]{
\parbox{\textwidth}{
Thomas Barclay$^{1,2}$\thanks{E-mail: tsb@arm.ac.uk}, 
Gavin Ramsay$^{1}$, 
Pasi Hakala$^{3}$,
Ralf Napiwotzki$^{4}$,
Gijs Nelemans$^{5}$,
Stephen Potter$^{6}$,
Ian Todd$^{7}$
}
\vspace*{4pt}\\
$^{1}$ Armagh Observatory, College Hill, Armagh, BT61 9DG, Northern Ireland, 
UK\\ 
$^{2}$ Mullard Space Science Laboratory, University College London,
Holmbury St. Mary, Dorking, Surrey, RH5 6NT, England, UK\\
$^{3}$Finnish Centre for Astronomy with ESO, University of Turku, V\"{a}is\"{a}l\"{a}ntie 20, FI-21500 PIIKKI\"{O}, Finland\\
$^{4}$ Centre for Astrophysics Research, University of Hertfordshire, College Lane, Hatfield AL10 9AB, UK\\
$^{5}$ Department of Astrophysics/IMAPP, Radboud University Nijmegen, PO Box 9010, NL-6500 GL Nijmegen, The Netherlands\\
$^{6}$ South African Astronomical Observatory, PO Box 9, Observatory 7935, Cape Town, South Africa\\
$^{7}$ Astrophysics Research Centre, School of Mathematics \& Physics, Queen's University, University Road, Belfast BT7 1NN
}

\date{Accepted 2011 January 12.  Received 2011 January 12; in original form 2010 November 15}

\begin{document} 
\date{\today} 
\maketitle

\maketitle
\begin{abstract}
The Rapid Temporal Survey (RATS) explores the faint, variable sky. Our observations search a parameter space which, until now, has never been exploited from the ground. Our strategy involves observing the sky close to the Galactic plane with wide-field CCD cameras. An exposure is obtained approximately every minute with the total observation of each field lasting around 2 hours. In this paper we present the first 6 epochs of observations which were taken over 5 years from 2003--2008 and cover over 31 square degrees of which 16.2 is within $10^{\circ}$ of the Galactic plane. The number of stars contained in these data is over $3.0\times10^{6}$. We have developed a method of combining the output of two variability tests in order to detect variability on time-scales ranging from a few minutes to a few hours. Using this technique we find $1.2\times 10^{5}$ variables -- equal to 4.1 per cent of stars in our data. Follow-up spectroscopic observations have allowed us to identify the nature of a fraction of these sources. These include a pulsating white dwarf which appears to have a hot companion, a number of stars with A-type spectra that vary on a period in the range 20--35 min. Our primary goal is the discovery of new AM CVn systems: we find 66 sources which appear to show periodic modulation on a time-scales less than 40 min and a colour consistent with the known AM CVn systems. Of those sources for which we have spectra of, none appears to be an AM CVn system, although  we have 12 candidate AM CVn systems with periods less than 25 min for which spectra are still required. Although our numbers are not strongly constraining, they are consistent with the predictions of \citeauthor{nelemans01}

\end{abstract}

\begin{keywords}
surveys -- stars: variables: other -- Galaxy: stellar content -- methods: data analysis -- techniques: photometric
\end{keywords}

\section{Introduction}

In recent years much progress has been made in increasing our knowledge of the variable sky. The advent of wide-field CCDs has brought with it a new parameter space that is only now beginning to be exploited. Variability over the course of days to weeks is well served for by surveys, such as, Pan-STARRS \citep{kaiser02}. On shorter time-scales a few experiments, such as SuperWASP \citep{pollacco06}, are able to detect variability on time-scales as short as a few minutes. However, these wide-angle experiments are unable to reach stars fainter than $g'\ge15$. The Rapid Temporal Survey (RATS) addresses this by using wide-field cameras on 2-m class telescopes to detect short-period variability in stars as faint as $g'=23$.

%However, few experiments are concentrating on observing variability at much shorter time-scales and to depths of $g'\ge15$. Indeed, the parameter space encompassing variability on times-scales of a few minutes to tens of minutes has been virtually untapped. The Rapid Temporal Survey (RATS) addresses this.

One class of object that is known to vary on time-scales shorter than a few 10's of minutes are the AM CVn binaries. These systems have orbital periods shorter than 70 min and
spectra practically devoid of hydrogen \citep[see][for reviews]{nelemans05proc,solheim10}.
They are composed of white dwarfs accreting from the
hydrogen exhausted cores of their degenerate companions. They are predicted to be
the strongest {\sl known} sources of gravitational wave (GW) radiation
in the sky \citep[e.g.][]{stroeer06,roelofs07,roelofs10}. Further, they are amongst a small number of objects which future gravitational wave observatories will be able to study in detail and for which extensive complementary electromagnetic observations exist.
 
Modest progress has been made in discovering more AM CVn systems in
recent years -- there are currently 25 known systems. Those which
have been discovered recently have been identified using spectroscopic
data from the SDSS archive
\citep[e.g.][]{roelofs05,anderson05,anderson08,rau10}. However, {\sl all}
the SDSS sources have orbital periods in the range 25--70 min.
Systems with periods shorter than this are predicted to have higher
mass transfer rates and be stronger gravitational wave sources.

AM CVn systems with orbital periods less than approximately 40 min show peak-to-peak intensity variations of between 0.01 \citep[V803 Cen,][]{kepler87} and 0.30 mag \citep[HM Cnc,][]{ramsay02} on time-scales close to their orbital period. One route to the discovery of such systems is through deep, wide-field, high-cadence photometric surveys. Our survey, RATS, is currently the only ground-based survey which samples this parameter space. The Kepler satellite is able to observe a similar parameter space \citep{koch10}. However, the targets for which data are downloaded are predefined (currently high cadence data are obtained for %RN2 is -> for
around 500 sources), and as such are likely to miss the majority of sources which vary on short time-scales.

We outlined our strategy and our initial results from our first epoch
of observations, which were obtained using the Isaac Newton Telescope
(INT) on La Palma in Nov 2003, in \citet{ramsay05}. Since then we have
obtained data from an additional four %RN3a 3 -> four
epochs using the INT and one %RN3b 1 -> one
epoch
using the MPG/ESO 2.2m telescope on La Silla Observatory, Chile. Additionally, we
have made significant revisions to our data reduction procedure. 

In this paper we provide an overview of our initial results from data
covering 6 separate epochs and outline our new reduction procedure. The emphasis of this work is on sources which have
been found to show an intensity modulation on periods $\le$40 min and $\le25$ min in particular. We make a preliminary
estimate of the space density of AM CVn systems based on the results
of the observations. Future papers will focus
on sources with longer periods, such as contact and eclipsing binaries, as well
as non-periodic variability such as flare stars.

\section{Observations and image reduction}
\label{sec:obs}
\begin{figure}
\includegraphics{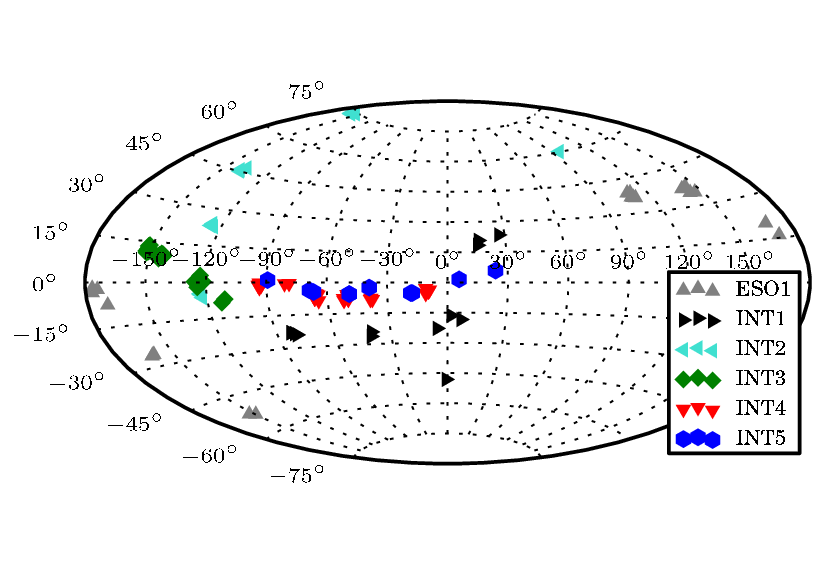}
\caption{The position of the field centres of all the fields observed during the first five years of the RATS project. The fields are plotted in Galactic coordinates using an Aitoff projection. Many of the fields are spatially close and so appear as only a single point in this figure.}
\label{fig:gal-proj}
\end{figure}

\begin{figure}
\includegraphics{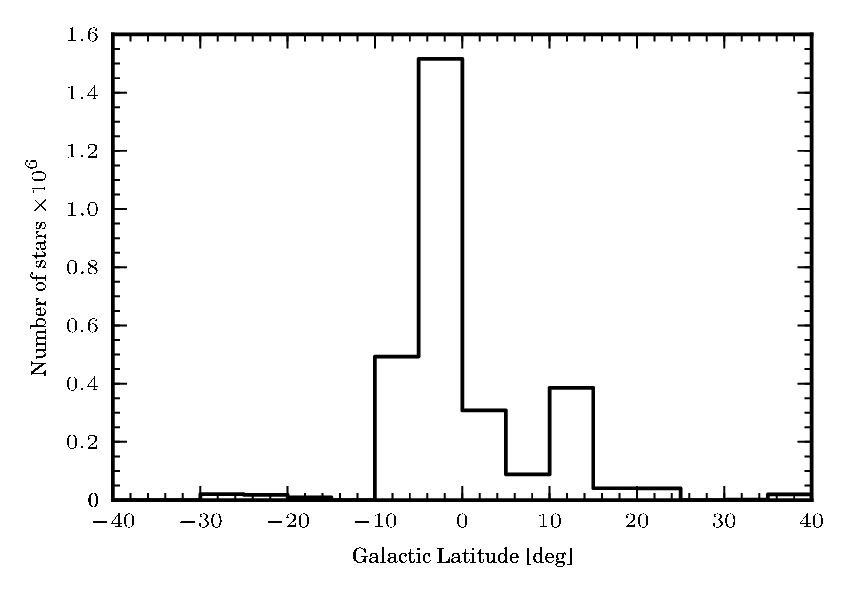}
\caption{The galactic latitudes of all the stars observed during the first five%RN6b 5->five
 years of the RATS project that have more than 60 photometric data points. Our fields are biased towards the Galactic equator: 78 percent of the stars in our sample lie within $|b|<10^{\circ}$.}
\label{fig:gallat-stars}
\end{figure}

Our data were taken at six separate epochs: five using the Wide Field Camera (WFC) on the INT and one using the Wide Field Imager (WFI) on the MPG/ESO 2.2m telescope (see Table~\ref{tab:obs} for details).
In our first epoch of observations, the fields were located at Galactic
latitudes with 20$^{\circ}<|b|<30^{\circ}$. Since then our fields have
been biased towards $|b|<15^{\circ}$ (see Fig.~\ref{fig:gal-proj} and \ref{fig:gallat-stars}). All fields are selected in such a way that no stars
brighter than $g\sim$12 are present. In addition, fields are typically chosen to be
close to the zenith to reduce differential atmospheric diffraction.

As during our first epoch of observations, we take a series of 30 second
exposures of the same field for $\sim$2 hours. With a dead-time
of $\sim$30 s for the INT observations and $\sim$110 s for the ESO/MPG 2.2-m fields, this results in approximately one observation every minute and every 2 minutes for the INT and ESO/MPG 2.2-m fields, respectively. To ensure
the highest possible signal-to-noise we do not use a filter.

Before the white light sequence commences we observe the field in a
number of filters. At earlier epochs we used Bessell \emph{B} and \emph{V} and either Bessell $I$ or SDSS $i'$ filters whilst at later epochs we used RGO
\emph{U} and SDSS $g'$ and $r'$ filters, with the addition of
a He\textsc{ii} 4686 \AA\ narrow-band filter at the latest epoch
(cf. Table~\ref{tab:obs}). For the INT2 and ESO1 observations an autoguider was used, for the other epochs no autoguider was used.

%RN6 changed 6,5,1->six,five,one
%RN5 changed this to what is below, I can't find a plot of QE for the 2.2m WFI:
%Our data were taken at six separate epochs: five using the Wide Field Camera (WFC) on the INT and one using the Wide Field Imager (WFI) on the MPG/ESO 2.2m telescope (cf. Tab.~\ref{tab:obs}). The WFC on the INT consists of four $4096\times2048$ pixel CCDs and has a total field of view of 0.28 square degrees, the WFI on the MPG/ESO 2.2m has eight $4098\times4046$ CCDs and a total field of view of 0.29 square degrees. We have observed a total of 110 fields which cover 31.3 square degrees of which 16.2 square degrees are at low galactic latitudes ($|b|< 10^{\circ}$).

 The WFC on the INT consists of four $4096\times2048$ pixel CCDs and has a total field of view of 0.28 square degrees. The four CCDs have a quantum efficiency which peak at 4600 \AA\ and an efficiency above 50 per cent from 3500--8000 \AA. The WFI on the MPG/ESO 2.2m has eight $4098\times4046$ CCDs which are optimised for sensitivity in the blue and has a total field of view of 0.29 square degrees. We have observed a total of 110 fields which cover 31.3 square degrees of which 16.2 square degrees are at low galactic latitudes ($|b|< 10^{\circ}$).

Images were bias subtracted and a flat-field was removed using twilight sky-images in the usual manner. In the case of our white light, $I$ and $i'$ band images, fringing was present (in the absence of thin cloud). We removed this effect by dividing by a fringe-map made using blank fields observed during the course of the night.

\begin{table}
\caption{Summary of the six %RN6a 6->six
 epochs in which our
 observations were taken. The number of stars column refers to sources with at least 60 photometric data points.}
\label{tab:obs}
\begin{tabular}{llrcrl}
\hline
Epoch&Dates&\# of&Galactic&Total&Filters\\
ID& &fields&latitudes& stars \\
\hline
INT1&20031128-30&12&$>|16^{\circ}|$&45572&$BVi'$\\
INT2&20050528-31&14&a&234029&$BVi'$\\
ESO1&20050603-07&20&b&750109&$BVI$\\
INT3&20070612-20&26&$<|15^{\circ}|$&1223803&$Ug'r'$\\
INT4&20071013-20&29&$<|10^{\circ}|$&678025&$Ug'r'$\\
INT5&20081103-09&9&$<|10^{\circ}|$&112788&$Ug'r'$+He\textsc{ii}\\
\hline
\multicolumn{6}{|l|}{
$^{a}$3 fields $<|10^{\circ}|$ and 11 fields $>|22^{\circ}|$}\\
\multicolumn{6}{|l|}{
$^{b}$4 fields $<|10^{\circ}|$ and 16 fields $>|16^{\circ}|$}
\end{tabular}
\end{table}

\section{Photometry}
\label{sec:phot}

\subsection{Extracting light curves}

In our first set of observations \citep{ramsay05} we used the \textsc{Starlink}\footnote{\textsc{Starlink} software and documentation can be obtained from http://starlink.jach.hawaii.edu/} aperture photometry package,
\textsc{Autophotom}. This technique provides accurate photometry at a reasonable computational speed for fields with low stellar density, but proved to be unsuitable for fields containing more than a few 1000 stars due to the computational processing time involved and -- for very crowded fields -- its inability to separate blended stars. 

For this reason we now use a modified version of \textsc{Dandia} \citep{bond01,bramich08,todd05} -- an implementation of difference image analysis \citep{alard98} -- which is more suited to crowded fields and takes into account changes in seeing conditions over the course of the observation. The source detection threshold is set to $3\sigma$ above the
background. We split each CCD into 8 sub-frames and calculated the point-spread function (PSF) for each sub-frame using stars that are a minimum of $14\sigma$ above the background and have no bad pixels nearby. A maximum of 22 stars are used in calculating the PSF. We use the four images with the best seeing in each field to create a reference frame. For each individual frame we degrade the reference frame to the PSF of that image and subtract the degraded reference frame. After subtraction we perform aperture photometry on the residuals. We do this for every frame and create a light curve for every star made up of positive and negative residuals.

As expected, our data suffer from systematic trends caused by effects such as
changes in airmass and variation in seeing and transparency \citep[for a discussion of systematic effects in wide-field surveys see][]{collier06} which can cause the spurious detections of
variable stars at specific periods. These periods are typically half
the observation length, although they can occur at other periods and are
field dependent. In order to minimise the effects of these trends we apply the
\textsc{Sysrem} algorithm \citep{tamuz05}. %RN8 added sentance on sysrem
The \textsc{Sysrem} algorithm assumes that systematic trends are correlated in a way analogous to colour-dependent atmospheric extinction, which is a function of airmass and the colour of each source. The colour is unique to each light curve and the airmass to each individual image though it does not necessarily refer to the true airmass but any linear systematic trend. These terms are minimised globally -- and the trend removed -- by modifying the measured brightness of each data point. We de-trend each CCD individually and use the method described in \citet{tamuz06} for running a variable number of cycles of the algorithm depending on the number of sources of systematic noise in the data, though we run a maximum of six cycles as we find that more than this starts to noticeably degrade signals in high-amplitude variables.

\begin{figure}
\includegraphics{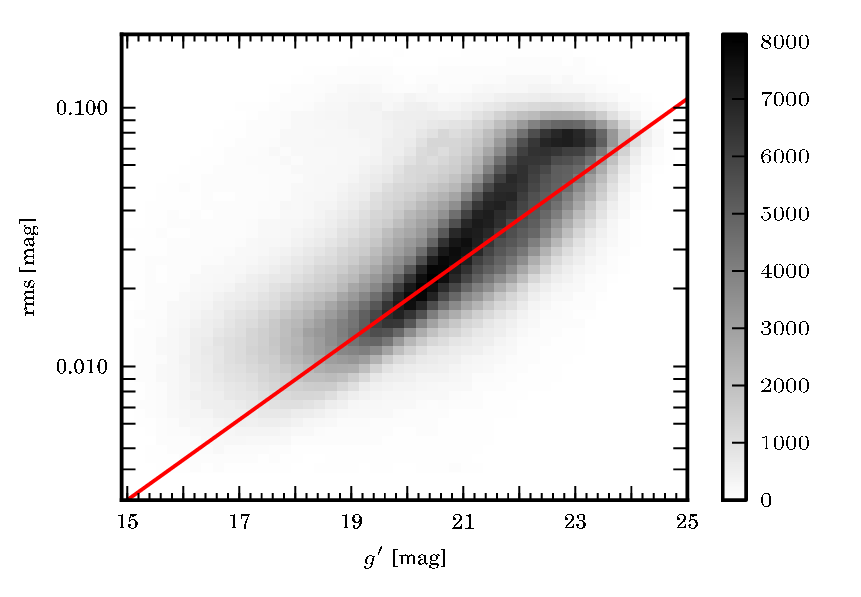}
\caption{The rms noise of each light curve plotted against $g'$ magnitude.
The grey-scale refers to the number of sources in each bin. The red line is the best-fit exponential function to the expected rms -- that is, the mean error of each light curve}
\label{fig:rms}
\end{figure}

To determine the quality of the resulting light curves we calculated
the root mean square (rms) from the mean for each light curve. When calculating the
rms we sigma-clip each light curve at the $5\sigma$ level in order to remove the effects of, say, single spurious data points. In Fig.~\ref{fig:rms} the measured rms is shown as a function
of the $g'$ mag for all stars in our sample. The mean rms of all the data is
0.046 mag with sources brighter and fainter than $g'=21.0$ having a mean
rms of 0.024 and 0.051 mag, respectively. If we look at the mean rms of
each field individually, we find two fields in our whole data-set with mean rms outside of 3
standard deviations which we attribute to very large variations in atmospheric transparency during
these observations. We show the best-fitting exponential function to the expected rms (equivalent to the average error on each light curve) in Fig.~\ref{fig:rms} and find it to be consistent with the measured rms except of the very faintest stars ($g'>22$).

\subsection{Determining colours}
When conditions appeared photometric we obtained images in different
filters of a number of Landolt standard fields \citep{landolt92}. {We
made use of data kindly supplied by www.astro-wise.org who give the
magnitude of Landolt stars in a range of different filters}. We
assumed the mean atmospheric extinction co-efficients for the
appropriate observing site. The resulting zero-points were very
similar to that
expected\footnote{e.g www.ast.cam.ac.uk/$\sim$wfcsur/technical/photom/zeros/}.

For our target fields we initially used \textsc{SExtractor}
\citep{bertin96} to obtain the magnitude of each star in each filter. 
However, in comparison to
\textsc{Daophot} \citep{stetson87}, \textsc{SExtractor} gave
systematically fainter magnitudes for faint sources. Since the
photometric zero-point for \textsc{Daophot} is derived from the PSF
(and therefore different from field-to-field) we calculated an offset
between the magnitudes of brighter stars determined using
\textsc{SExtractor} and \textsc{Daophot}. We then applied this 
offset to the magnitudes derived using \textsc{Daophot}. To convert
our $BVi'$ data (Table~\ref{tab:obs}) to $g'r'$ magnitudes we used the
transformation equations of \citet{jester05}. 

Although our light curves were obtained in white light we note the
depth of our observations as implied in the $g'$ filter. For stars with
$(g-r)\sim$1.0, the typical depth for fields observed in photometric
conditions and with reasonable seeing (better than 1.2 arcsec) is
$g\sim22.8-23.0$, while for redder stars ($g-r\sim$2.0) the depth is
$g\sim23.6-24.0$.

To test the accuracy of our resulting photometry, we obtained a small
number of images of SDSS fields \citep{york00}. For stars $g'<20$ we
found that for $g_{\mathrm{RATS}}-g_{\mathrm{SDSS}}$, $\sigma$=0.12 mag
and
$(g_{\mathrm{RATS}}-r_{\mathrm{RATS}})-(g_{\mathrm{SDSS}}-r_{\mathrm{SDSS}})$,
$\sigma$=0.22. For stars
$20<g<22$ we find for $g_{\mathrm{RATS}}-g_{\mathrm{SDSS}}$, $\sigma$=0.29
mag and
$(g_{RaTS}-r_{RATS})-(g_{\mathrm{SDSS}}-r_{\mathrm{SDSS}})$,
$\sigma$=0.27. Given our
project is not optimised to achieve especially accurate photometry
these tests show that our photometric accuracy is sufficient for our
purposes, namely determining an objects brightness and approximate colour.

\subsection{Astrometry}

As part of our pipeline we embedded sky co-ordinates into our images
using software made available by Astrometry.net \citep{lang10}. This
uses a cleaned version of the USNO-B catalogue \citep{barron08} as a
template for matching sources in the given field. The only input we
provide is the scale for the detectors and the approximate position of
the field, which is taken from the header information in the
images. The
Astrometry.net software works well in either sparsely or relatively
dense fields. By comparing the resulting sky co-ordinates of stars
with matching sources in the 2MASS \citep{jarrett00} the typical error
was 0.3--0.5 arcsec.

\section{Variability}
\label{variable}

Due to the large data-set, it is necessary for us to automate the
detection of variable sources. We find that no single algorithm is
appropriate for the detection of all types of variable sources
present in our data and in most cases for the detection of even a
single class of variable source since the false positives are unacceptably high if we use just one algorithm. We therefore, typically use at least two independent algorithms to detect each class of variable object.

Before passing the light curve data to the variability detection
algorithms we remove light curves which contain less than 60 data
points. Of the initial 3.7 million stars, this leaves 3.0 million. We
remove light curves with relatively few data points because our
variability detection algorithms can produce spurious results when
a significant amount of data are missing. This process prevents
the discovery of transient phenomena, an aspect which we will
investigate in more detail in the future.

In future work we will discuss sources such as contact and eclipsing binaries and flare stars whose variability is not periodic over a two hour time-frame. However, in this paper we will concentrate on the detection of periodic variables.

\subsection{Periodic variability detection}
\label{sec:per-var}

We use two algorithms in order to identify variable sources: analysis of variance (AoV), and the
Lomb-Scargle periodogram (LS). From these algorithms we determine the
analysis of variance formal false-alarm probability, AoV-FAP; and the Lomb-Scargle formal false-alarm
probability, LS-FAP. We use the \textsc{Vartools} suite of software to
calculate these parameters \citep{hartman08}.

\begin{figure}
\centering
\includegraphics{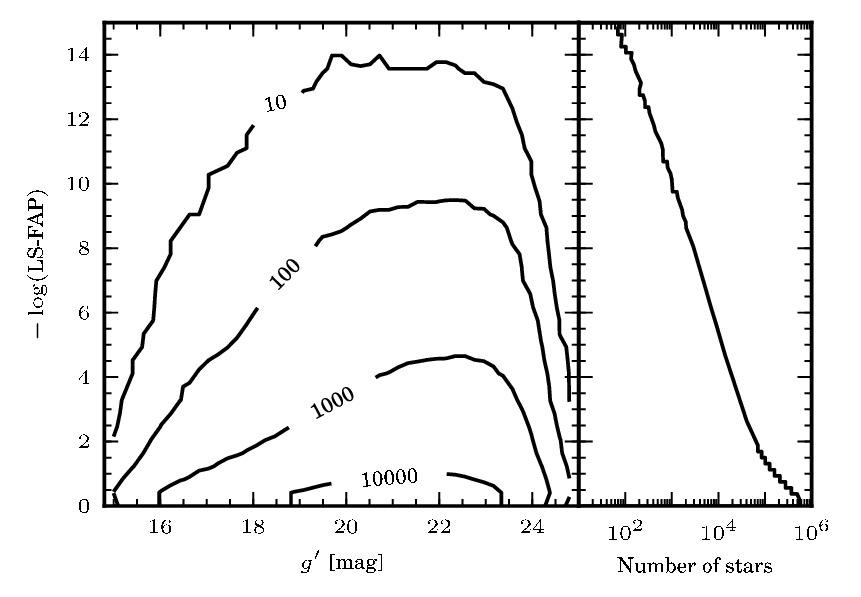}
\caption{The distribution of the LS-FAP statistic in magnitude and number where the LS-FAP is for the highest peak in the frequency range. The contours refer to the number of stars in each bin where the data was been binned and has a bin size of 0.2 and 0.4 in magnitude and $\log(\textrm{LS-FAP})$, respectively.}%RN16 added highest peak..
\label{fig:lsfap}
\end{figure}

The Lomb-Scargle periodogram \citep{lomb76,scargle82,press89,press92} is
an algorithm designed to pick out periodic variables in unevenly sampled data.
As a test of variability we use the LS-FAP. Its distribution as a function of magnitude is shown in Fig.~\ref{fig:lsfap}.
This parameter is a measure of the probability that the highest peak in the
periodogram is due to random noise. If the noise in our data were frequency
independent the LS-FAP would refer to the probability of the detected period
being due to random noise. However, our data are subject to sources of systematic error which we attribute to red noise: these include the number of data points in the light curve and the range
in airmass at which a star is observed. Hence, we use it as a relative measure of variability.

\begin{figure}
\centering
\includegraphics{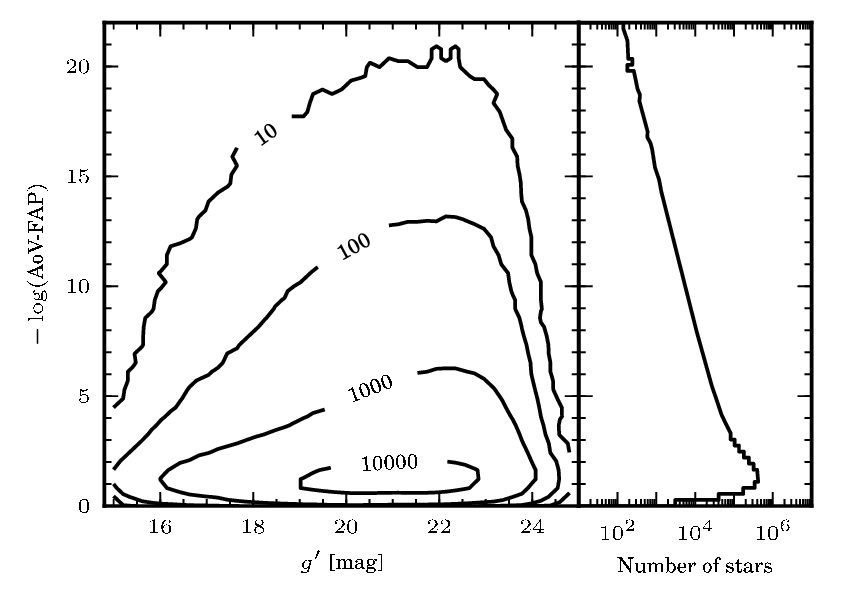}
\caption{A similar plot to Fig.~\ref{fig:lsfap} but this time showing the distributions for AoV-FAP. The data is binned and the bin size is 0.2 and 0.4 in magnitude and $\log(\textrm{AoV-FAP})$, respectively.}
\label{fig:aovfap}
\end{figure}

We use a modified implementation of the analysis of variance
periodogram \citep{schwarzenberg89,devor05}. The AoV algorithm folds the light curve and selects the period which minimises the variances of a second-order polynomial in eight phase-bins. A periodic
variable will have a small scatter around its intrinsic period and high scatter
on all other periods. The statistic $\Theta_{AoV}$ is a measure of the goodness of the fit to the best fitting period, with larger values indicating a better fit. In order to be consistent with the LS-FAP we calculate the formal false-alarm probability of the detected period being due to random noise (AoV-FAP) -- with the same caveats as with the LS-FAP -- using the method described by \citet{horne86}. We show the distribution of the AoV-FAP statistic in number and as a function of magnitude in Fig.~\ref{fig:aovfap}.

The AoV algorithm, while similar to the LS method, should allow better
variable detection as it fits a constant term to the data as opposed
to subtracting from the mean as is done in the LS routine
\citep{hartman08}. However we find that AoV has a number of negative features. It suffers from severe aliasing at periods of 2--3 min and for this reason we only search for periods longer than 4 min. In addition there is a tendency to detect a multiple of the true period when the true period is less than $\sim$40 min. In tests with simulated light curves we found approximately 10 per cent of sources with a period of 20 min were detected by AoV as having 40 min periods. The main weakness of the LS algorithm is that if a source has a periodic modulation in brightness for only a small amount of the total light curve then -- according to the LS-FAP -- it is detected as significantly variable. This leads to a large number of false-positive detections which are probably due to random noise. To combat this we have developed a technique to combine the AoV and LS algorithms.

\begin{figure*}
\includegraphics{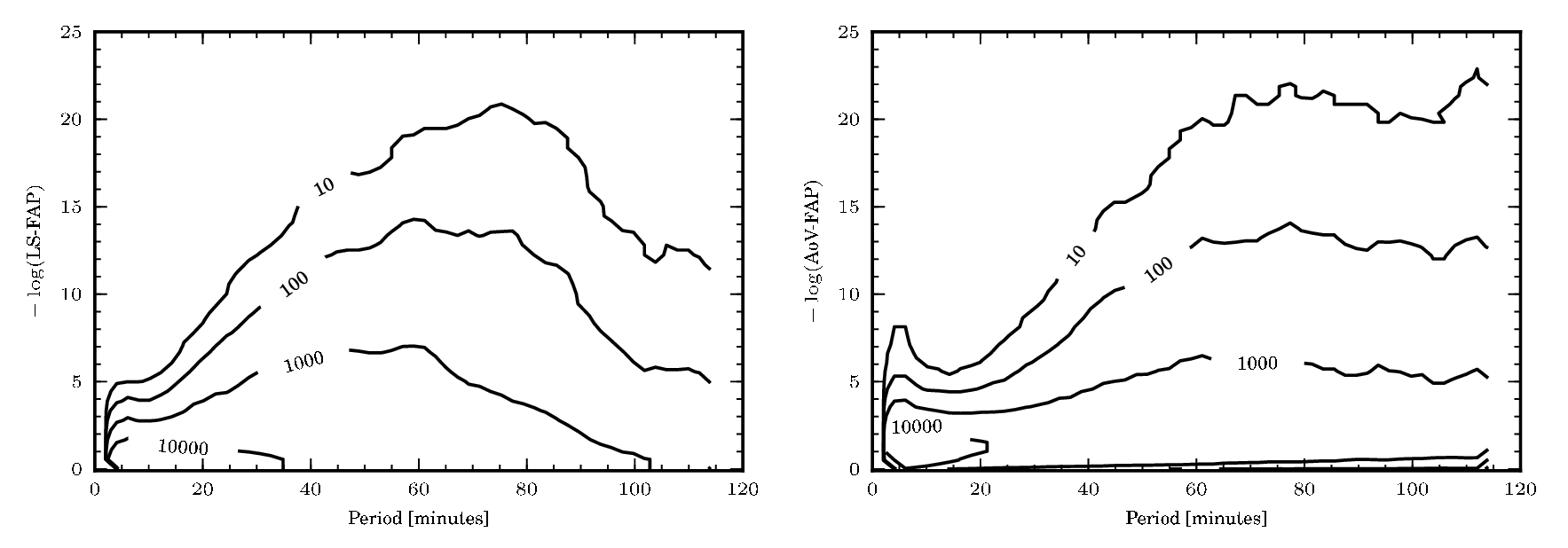}
\caption{LS-FAP and AoV-FAP statistics in the left and right plots respectively,
plotted against the period measured by those statistics. Spurious detections of
variability are obvious at very short periods in AoV-FAP, while LS-FAP is less sensitive to longer period variability.}
\label{fig:aov-fap-per}
\end{figure*}

Our technique, which combines the LS-FAP and AoV-FAP statistics is a multi-stage process. The first step is to determine if the source is detected as significantly variable by both the LS and AoV algorithms: we then test whether the period each algorithm detects is the same. Shown in Fig.~\ref{fig:aov-fap-per} are the AoV-FAP and LS-FAP statistics plotted against the period that is measured by the respective
methods. We can see here that both algorithms suffer from deficiencies: the distributions of AoV-FAP and LS-FAP are not constant
with period, but tend to
higher significance at longer periods. In order to account for
the bias of the distribution we use an approach whereby we bin the data in
period with each bin 2 min wide. A source passes the first two stages of the algorithm if it is above a specific significance in both AoV-FAP and LS-FAP relative to the other sources in the period bin. In order to determine this significance we use the median absolute deviation from the median \citep[MAD,][]{hampel74} which is defined for batch of parameters $\{x_{1},\ldots,x_{n}\}$ as
\begin{equation}
 \mathrm{MAD}_{n} = b\: \mathrm{med}_{i} |x_{i} - \mathrm{med}_{j}x_{j}|
\end{equation}
where $b$ is a constant which makes the parameter consistent with the standard deviation. For a Gaussian distribution $b=1.4826$ \citep{rousseeuw93} which we use for simplicity. We use the median, as using the mean is not appropriate when the first moment of the distribution tail is large \citep{press92}; the large tails in the distributions of AoV-FAP and LS-FAP are shown in the right-hand plots in Figs.~\ref{fig:lsfap} and \ref{fig:aovfap}. The median and MAD are more robust statistics.

We vary the number of MADs a source must be above the median to be detected depending on epoch as the distributions of LS-FAP and AoV-FAP parameters are different. For INT2 and ESO1 we use 200 MADs, for INT1, INT4 and INT5 we use 800 MADs. These number of MADs above the median are used as they provide an appropriate balance between low amplitude detections and false positives -- which we discuss in \S~\ref{sec:falsealarm} and \S~\ref{sec:fap-aov} -- we attribute the need for different numbers of MAD above the median to the use of an autoguider on INT2 and ESO1 and not on the other epochs. Due to different epochs having different distributions of variability parameters we calculate the median independently for each epoch.

Both AoV and LS algorithms produce a periodogram; from the highest peak in the periodogram we calculate the most likely period of a given light curve. We take all the candidate
variables and test whether the period detected by AoV matches that
detected by LS. We class a period as a match if 
\begin{equation}
P_{AoV} \pm \Delta P_{AoV} = P_{LS} \mp \Delta P_{LS}
\label{eq:PdelP}
\end{equation}
where $P_{AoV}$ and $P_{LS}$ are periods detected by the two algorithms AoV and LS, respectively. $\Delta P$ is the error in measured period. We determine $\Delta P$ using an approximation of equation (25) in \citet{schwarzenberg91} whereby we assume that 
\begin{equation}
\Delta P/P^{2} \sim k.
\label{eq:delP}
\end{equation}
 This assumption holds for all but the lowest signal-to-noise detection of variability. In order to determine an appropriate value for the constant, k, we inject sinusoidal signals of various periods into non-variable light curves and measure the standard deviation on $|P_{AoV} - P_{LS}|$. We set the constant, k, in Eq.~\ref{eq:delP} so as to give a $\Delta P$ at a given period equal to twice the standard deviation of $|P_{AoV} - P_{LS}|$. We find $k=0.002$ to be appropriate. The AoV algorithm has an annoying habit of detecting a multiple of the true period, so for this reason we modify Eq.~\ref{eq:PdelP} to
\begin{equation}
\frac{P_{AoV}}{n} \pm \Delta P_{AoV} = P_{LS} \mp \Delta P_{LS}
\label{eq:PdelP2}
\end{equation}
where $n = \{1,2,3,4\}$.

\begin{figure}
\includegraphics{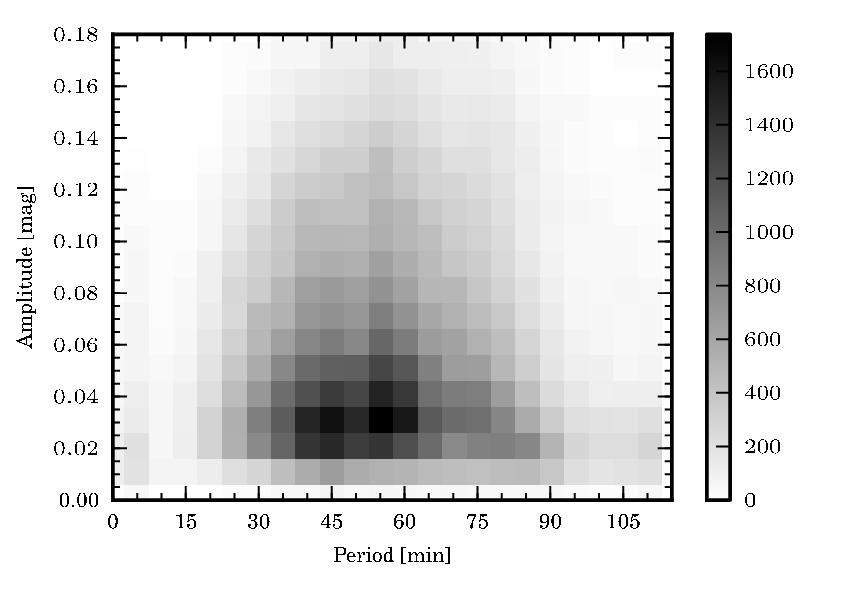}
\caption{The amplitude and period of sources classes as variable using the technique for combining LS-FAP and AoV-FAP. The grey-scale uses bin size of 0.01 mag in amplitude and 5 min in period and the colour refers to the number of sources in that bin.}
\label{fig:perdist}
\end{figure}

Sources that have matching periods and have been classified as candidate variable sources by both AoV-FAP and LS-FAP are then regarded as 'significantly' variable sources. We detect 124334 stars which show variability on a timescale of 4--115 min: the distribution of the measured periods are shown in Fig.~\ref{fig:perdist}. We caution that this technique can detect variables that are not truly periodic -- many flare stars have detected periods near the observation length -- or may have periods longer than that detected by our method -- contact binaries typically have a true period twice the measured one. If a period of less than half the observation length is measured then this is likely to be a true period. However, longer periods detected by the LS and AoV algorithms indicate only that the source varies significantly on time-scales less than $\sim$2 hours.

\subsection{False positives}
\label{sec:falsealarm}
In order to determine the false positive detection rate, that is, the
chance of a source with variability due to noise being identified as a
real variable, we pick a light curve at random from the whole
data set and construct a new light curve using a bootstrapping approach. The light curve consists of three columns of data: time, flux and error on the flux. We keep the time column as it is, and for a light curve with N individual photometric observations, randomly select 
N fluxes and errors from the N points in the original light curve. We do not limit the number of times a flux-error combination is selected. The reason for reconstructing a light curve in this fashion is that any periodic variability which is present in the original light curve is removed, allowing us to measure the chance that any variability present is due to noise.

We reconstruct $10^5$ bright
and $10^5$ faint randomly selected light curves and attempt to detect variability using our method for combining the AoV and LS algorithms. Bright and faint refers to
sources brighter than and fainter than $g'=21.0$, respectively where
21.0 is approximately the median $g'$ magnitude. For the bright
sample we class 10 stars as variable and for the faint sample this
increases to 17. This equates to a false positive rate of 0.01 and
0.02 per cent for bright and faint sources, respectively.

To find the improvement in the false positive detection rate we run the same routine but using only one of the statistics, i.e. LS-FAP or AoV-FAP. The method of detection is the same as the first stage of the two algorithm method -- is a source found to have a variability statistic above the detection threshold for its period. When using both AoV-FAP and LS-FAP the false-positive rate is around 0.5 per cent. When using the two algorithm method the number of false positives is reduced by a factor of $\sim40$.

\subsection{Sensitivity tests as a function of amplitude and 
period}
\label{sec:fap-aov}

To determine the space densities for different classes of sources which vary on time-scales of less than $\sim$2 h
it is essential that we determine our sensitivity to different brightnesses, periods and amplitudes. To do this we inject sinusoids of
known period and amplitude into non-variable light curves and then
attempt to detect it using our LS-FAP + AoV-FAP test. 
%In order to do this we split the data into two groups based on whether they are brighter or fainter than $g' = 21.0$ which is approximately the median $g'$ magnitude, and we refer to as the bright and the faint group, respectively. 
We split the sources into a bright and faint groups -- brighter or fainter than $g' = 21.0$. -- and for each brightness range we inject a periodic signal
into a non-variable light curve. The non-variable light curve is drawn
randomly from a pool of light curves that have AoV-FAP and LS-FAP
statistics within 0.5 median absolute deviations of the median AoV-FAP and LS-FAP
of all light curves with $g'$ greater than and less than 21.0 for
the bright and faint groups, respectively. The periodic signal
injected is drawn from a grid of period-amplitude combinations where
the periods range from $4-120$ min and amplitudes from $0.02-0.2$
mag. We define amplitude as peak-to-peak. The advantage of using real (non-variable) light curves over simulated data is
that it preserves the noise values which may be non-Gaussian.

\begin{figure*}
\includegraphics[width=\textwidth]{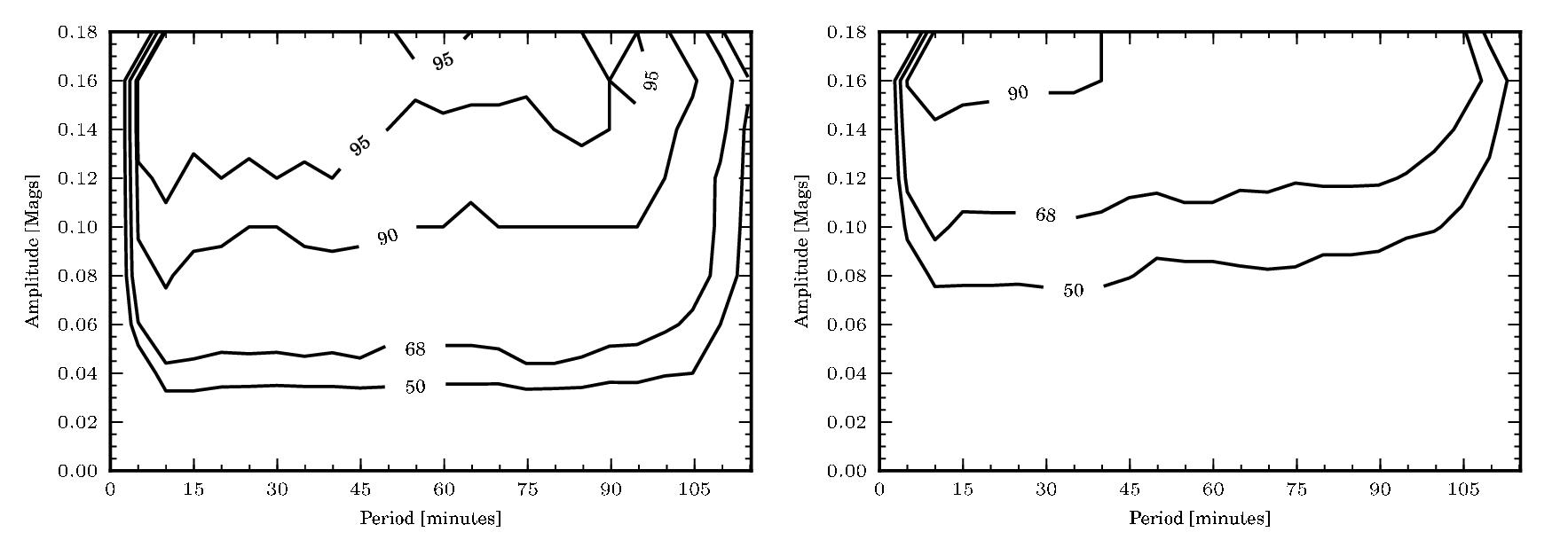}
\caption{Our sensitivity to periodic sources of different period and
peak-to-peak amplitudes. In the left plot, the contours refer to the percentage of sources
detected as variable when the original non-variable light curve was
brighter than $g' = 21.0$. The right plot show the same but for sources
fainter than $g'=21.0$.}
\label{fig:detection}
\end{figure*}

We run the entire grid 100 times for each brightness range which allows
us to build up confidence of a variable with a given period and
amplitude being detected. The results of this are plotted in
Fig.~\ref{fig:detection} and show that in the brighter sample, sources
with a period less than 90 min have a $\sim$70 per cent chance of being
detected if they have amplitudes larger than 0.05 mag, rising to
above a 90 per cent chance for amplitudes greater than $\sim0.10$ mag. Stars
fainter than $g'=21.0$ with an injected period less than 90 min have
$\sim$ 50 per cent detection chance if they have an amplitude greater than 0.08 mag and only sources with a period less than 40 min and an amplitude greater than $\sim$0.15 mag have a 90 per cent chance of detection.

From these results we find that our LS-FAP + AoV-FAP method is relatively
good at identifying variables with periods less than 90 min in bright sources but is weaker at identifying periodic variability in the fainter sources. The advantage of
this method is the small number of false-positives expected to be detected.

\subsection{Summary of variables in RATS}
We have detected 124334 stars which show variability on timescales 4--115 min. This equates to 4.1 per cent of the total stars -- with at least 60 data points -- in our survey, of which 22679 sources have periods detected by the LS algorithm of less than 40 min. We expect 0.02 per cent -- equal to $\sim$600 sources -- of the variables detected in the RATS data to be false positives. The median value of the $g'$ band magnitude is 21.0, whereas the median brightness of all variables is 21.5 and only 30 per cent of variables are brighter than the median brightness of all stars in the RATS data.

\section{Colour of sources which vary on time-scales of less than two hours}

\begin{figure*}
\includegraphics{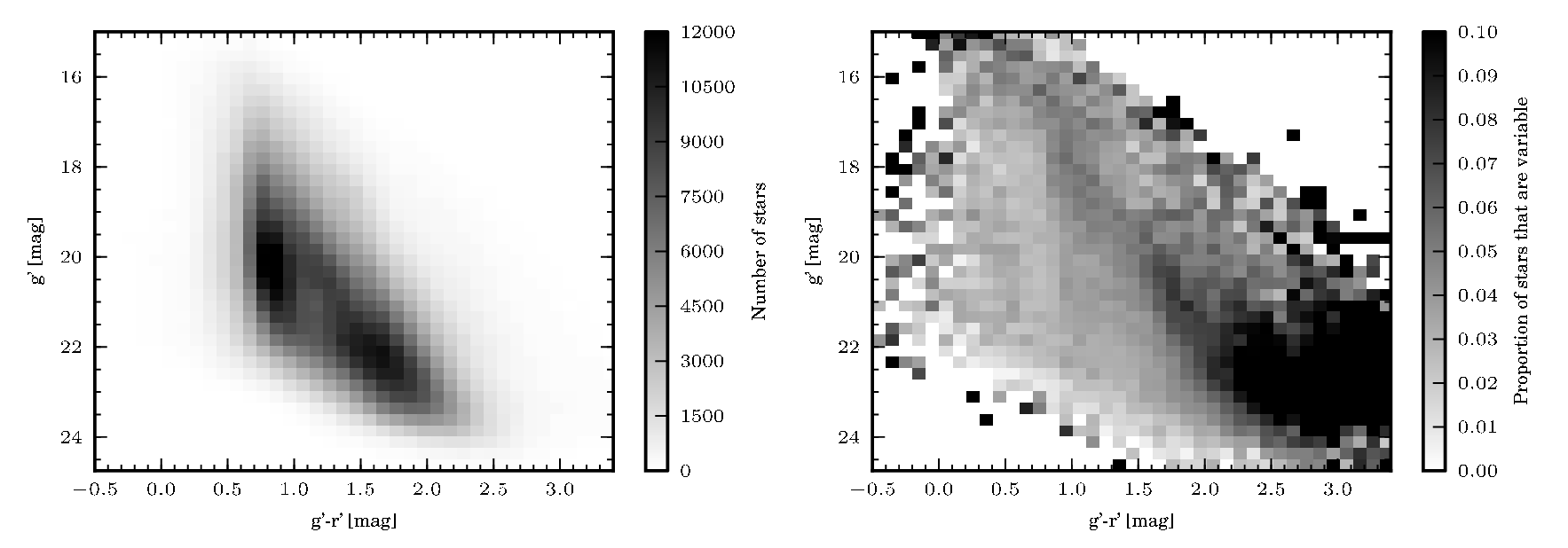}
\caption{Colour-magnitude diagram showing $g'-r'$ against $g'$
magnitude. The left plot contains all the stars in our data, the right plot show the proportion of stars in
each bin that are classified as variable using the method in \S\ref{sec:fap-aov}.}
\label{fig:cmd}
\end{figure*}

We show the colours of all the stars in our data in the $g'-r',
g'$ plane in left hand panel of Fig.~\ref{fig:cmd}. We note the
presence of two broad populations; one which is bright ($g'\sim19-21$) and blue
($g'-r'\sim0.6-0.7$), and one which is fainter ($g'\sim22-23$) and redder
($g'-r'\sim1.5-2.0$). The bluer population is thought to
originate in the Galactic halo or thick disk, while the redder
population is thought to originate in the thin disk
\citep[e.g.][]{robin03}. Data similar to these has been used to model
the structure of the Milky Way as a function of Galactic latitude and
longitude \citep[cf.][]{chen01}, but this is beyond the scope of
the present work.

We also show in Figure \ref{fig:cmd} the stars which have been
classified as variable (cf. \S\ref{variable}) in the $g'-r', g'$
plane. The variable sources are concentrated at the faint, red end of the of the colour-magnitude diagram. In contrast, the Faint
Sky Variability Survey (FSVS) found that the variable sources they
detected were biased towards bluer colours \citep{morales06}. We
attribute this difference to the fact that the FSVS was more typically
sensitive to longer time-scale variations than our survey, coupled with
the fact that they observed fields at mid to high Galactic latitudes
\citep{groot03}.

\section{Short period blue variables}
AM CVn systems show optical modulation on a period close to their orbital period and are intrinsically blue in colour -- the reddest have $g'-r'$ colours of $-0.1$ -- and modulate on periods of $<40$ min with amplitudes from 0.01--0.30 mag. In this section we restrict our search to stars with parameters fulfilling these criteria. 

The known AM CVn systems found in SDSS data have de-reddened $g'-r' < -0.1$ \citep{roelofs09}. The majority of our stars are close to the Galactic plane where the interstellar reddening is high. For fields close to the Galactic plane ($|b| < 15^{\circ}$) we adopt an average value for the total neutral hydrogen column density to the edge of the Galaxy of $N_{H}=3.67\times 10^{21}$ cm$^{-2}$ which is taken from the Leiden/Argentine/Bonn (LAB) Survey of Galactic HI \citep{kalberla05}. We calculate the optical extinction in the $V$ band, $A_{V}$ using the relation found by \citet{guver09}:
\begin{equation}
N_{H} = (2.21\pm 0.09)\times 10^{21} A_{V}.
\end{equation}
We take $A_{V} = 3.10\times E(B-V)$ \citep{fitzpatrick99} which gives $A_{V} = 1.66$ and $E(B-V) = 0.54$. This result is consistent with \citet{joshi05} who observed open clusters close to the Galactic plane ($|b| < 5^{\circ}$) and find $E(B-V) = 0.6$. If we convert this extinction calculated from the $N_{H}$ of our fields to Sloan $g'$ and $r'$ filters we find $E(g'-r') = 0.60$. Adding this to the red cutoff for AM CVn systems $(g'-r' <-0.1)$ we get a cutoff of for AM CVn systems in our data which are close to the Galactic plane, of $g'-r' < 0.5$

After these cuts the number of sources left as candidate AM CVn systems is 250. We visually inspect these as a final level of quality control, and class 66 of these short-period blue variables as `good' candidates -- that is, having a strong period and no obvious systematic effects present. 

\begin{figure}
\includegraphics{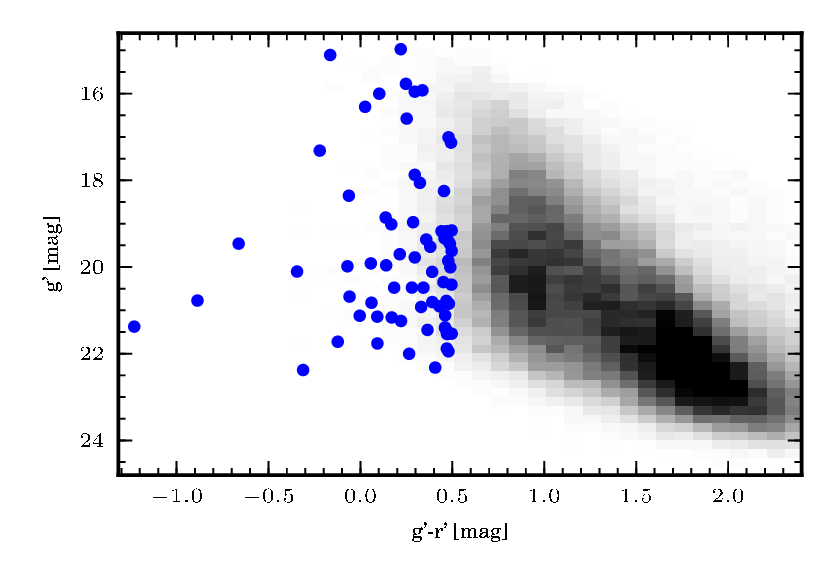}
\caption{Colour-magnitude diagram of all the detected variable stars plotted as
grey-scale. Those stars identified as short period ($P < 40$ min) blue
variables -- shown in Table~\ref{tab:bspv25} and \ref{tab:bspv40} -- are shown as blue dots.}
\label{fig:cmd-spv}
\end{figure}

\begin{figure*}
\includegraphics{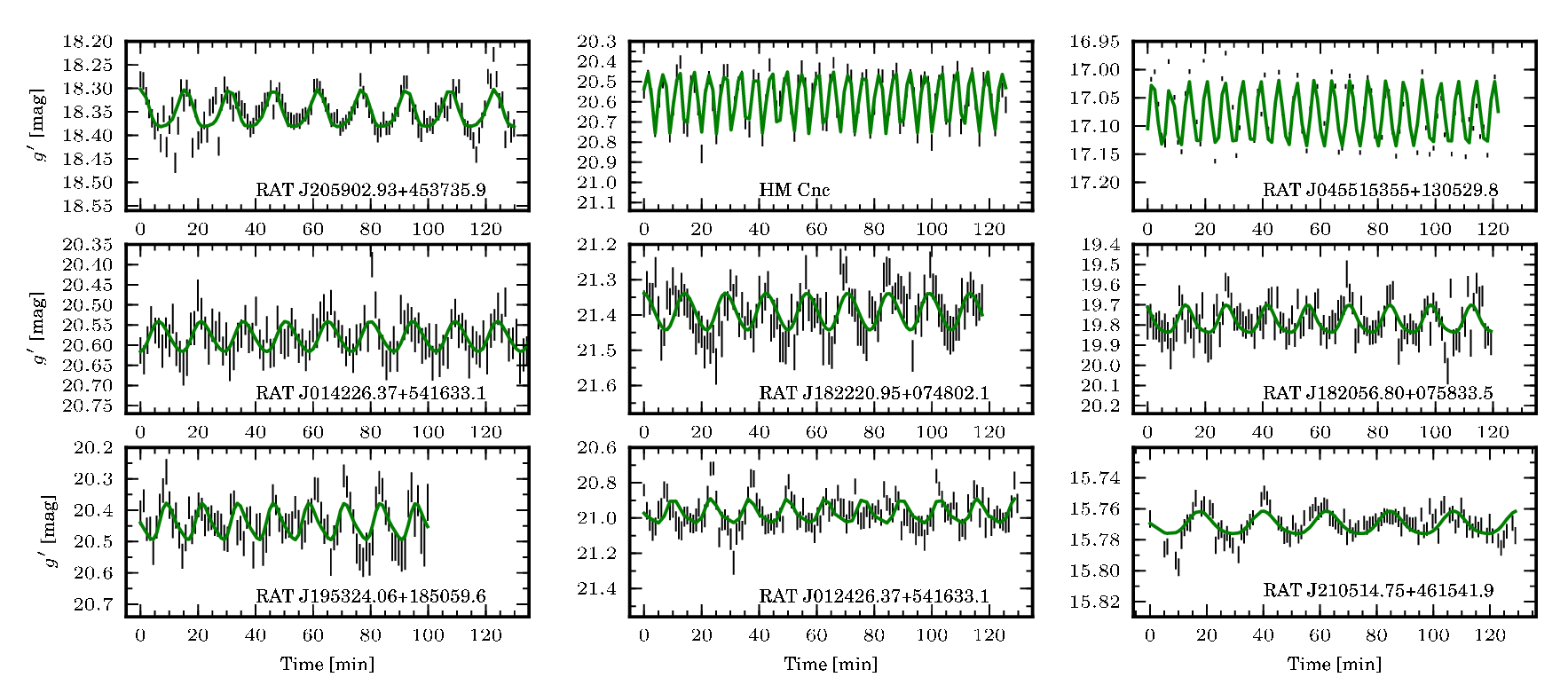}
\caption{A sample of the blue short-period variable stars shown in Table~\ref{tab:bspv25}. The green curve shows the best-fitting amplitude, period and phase of the light curve. The period was found using the Lomb-Scargle periodogram algorithm, the amplitude and phase from fitting a sine curve to the light curve on that period.}
\label{fig:bspv-lc}
\end{figure*}

 In Tabs.~\ref{tab:bspv25} and \ref{tab:bspv40} we list the short period blue variables for those sources with periods less than 25 min and for those with periods 25--40 min, respectively. We split these into two tables for ease of comparison with \citet{nelemans01,nelemans04} model of the Galactic population of AM CVn systems which is only relevant for sources with periods less than 25 min (see \S~\ref{sec:spacedensity}). The sources contained in these two tables are plotted as blue dots in Fig.~\ref{fig:cmd-spv}. A subset of the light curves of these sources are shown in Fig.~\ref{fig:bspv-lc}. We fit a sinusoid to each light curve at the period of the highest peak in the Lomb-Scargle periodogram and this is shown over-plotted on each light curve.

\begin{table*}
\caption{Sources identified as periodic variable stars with periods less than 25
min and a $g'-r'$ colour of less than 0.5. They have also passed a visual quality check. The period is determined from the Lomb-Scargle periodogram and the amplitude is peak-to-peak of the best fitting sine curve. If classification of the source has been possible then this is stated in the notes column. For the previously know sources the classification come from the literature, for the newly discovered variables we classify the sources based on spectral type and the spectral fitting described in \S~\ref{sec:fitting}. The Cat. name column is the unique identifier for that source and is of the format EpochID-Field\#-CCD\#-Star\#.}
\label{tab:bspv25}
\begin{tabular}{|l|l|l|r|r|r|r|l|l|}
\hline
  \multicolumn{1}{|c|}{Cat. name} &
  \multicolumn{1}{c|}{R.A. (J2000)} &
  \multicolumn{1}{c|}{Dec. (J2000)} &
  \multicolumn{1}{c|}{Period [min]} &
  \multicolumn{1}{c|}{$g'$ [mag]} &
  \multicolumn{1}{c|}{$g'-r'$ [mag]} &
  \multicolumn{1}{c|}{Amplitude [mag]} &
  \multicolumn{1}{c|}{Notes} &
  \multicolumn{1}{c|}{Ref.} \\
\hline
  4-7-3-9108 & 01:41:31.42 & +55:34:11.3 & 5.31 & 20.48 & 0.18 & 0.033 &  &3 \\
  4-11-1-2519 & 01:42:26.37 & +54:16:33.1 & 13.23 & 21.12 & $-$0.00 & 0.128 &1& 3\\
  5-6-3-4828 & 04:32:10.40 & +40:33:33.9 & 8.43 & 20.10 & $-$0.34 & 0.249 &1 &3 \\
  1-16-4-1316 & 04:55:15.22 & +13:05:29.8 & 6.24 & 17.32 & $-$0.22 & 0.115 &Pulsating sdB star  &4 \\
  4-27-3-662 & 05:07:51.28 & +34:31:55.5 & 22.15 & 16.31 & 0.03 & 0.029 &2  &3 \\
  1-22-4-1415 & 07:39:19.05 & +23:52:39.9 & 14.71 & 20.83 & 0.06 & 0.074 &1&1 \\
  1-806-4-713 & 08:06:22.95 & +15:27:31.2 & 5.36 & 20.77 & $-$0.89 & 0.284 &AM CVn star  &5 \\
  3-2-3-14617 & 17:58:46.34 & +01:45:59.8 & 9.58 & 22.00 & 0.26 & 0.103 &  &3 \\
  3-5-1-25551 & 17:59:52.90 & +01:22:19.4 & 22.10 & 20.01 & 0.49 & 0.028 &  &3 \\
  3-5-3-4143 & 18:00:57.72 & +01:42:03.0 & 20.40 & 20.48 & 0.34 & 0.038 &  &3 \\
  2-7-2-1301 & 18:01:52.91 & +29:42:30.4 & 8.55 & 19.92 & 0.06 & 0.058 &  &3 \\
  3-26-1-604 & 18:20:56.80 & +07:58:33.5 & 14.20 & 19.96 & 0.14 & 0.134 &  &3 \\
  3-27-2-6182 & 18:22:20.95 & +07:48:02.1 & 14.17 & 21.24 & 0.22 & 0.103 &1&3 \\
  2-11-4-34562 & 19:53:24.03 & +18:50:59.6 & 12.38 & 20.68 & $-$0.06 & 0.109 &  &3 \\
  2-11-3-25021 & 19:53:27.17 & +18:59:14.4 & 20.00 & 19.98 & $-$0.07 & 0.240 &Dwarf nova  &6 \\
  4-9-2-9694 & 20:58:00.24 & +45:32:32.8 & 7.23 & 19.16 & 0.50 & 0.039 &  &3 \\
  4-9-4-14918 & 20:59:02.93 & +45:37:35.9 & 15.34 & 18.36 & $-$0.06 & 0.080 &Pulsating white dwarf  &3 \\
  4-16-4-21281 & 21:05:14.75 & +46:15:41.9 & 22.42 & 15.77 & 0.25 & 0.014 &2 &3 \\
  1-3-2-46 & 23:03:11.95 & +34:31:04.1 & 9.20 & 20.78 & 0.47 & 0.040 &  &3 \\
  4-6-4-2813 & 23:49:06.89 & +56:24:40.4 & 22.73 & 18.06 & 0.32 & 0.039 &2 &3 \\
  4-6-3-470 & 23:49:29.00 & +56:34:26.2 & 5.82 & 16.58 & 0.25 & 0.018 &  &3 \\
  4-25-3-9719 & 23:52:42.07 & +56:31:53.2 & 12.85 & 19.46 & $-$0.66 & 0.098 &  &3 \\
\hline
\multicolumn{9}{|l|}{$^{1}$Low signal-to-noise spectrum -- no obvious emission lines,$^{2}$A-type star spectrum}\\
\multicolumn{9}{|l|}{
 $^{3}$This paper, $^{4}$\citet{ramsay06}, $^{5}$\citet{ramsay02}, $^{6}$\citet{ramsay09}}
\end{tabular}
\end{table*}

\begin{table*}
\caption{The same as Table~\ref{tab:bspv25} for sources with periods between 25 and 40 min.}
\label{tab:bspv40}
\begin{tabular}{|l|l|l|r|r|r|r|l|l|}
\hline
  \multicolumn{1}{|c|}{Cat. name} &
  \multicolumn{1}{c|}{R.A. (J2000)} &
  \multicolumn{1}{c|}{Dec. (J2000)} &
  \multicolumn{1}{c|}{Period [min]} &
  \multicolumn{1}{c|}{$g'$ [mag]} &
  \multicolumn{1}{c|}{$g'-r'$ [mag]} &
  \multicolumn{1}{c|}{Amplitude [mag]} &
  \multicolumn{1}{c|}{Notes} &
  \multicolumn{1}{c|}{Ref.} \\
\hline
 5-7-2-1129 & 06:54:32.71 & +10:31:52.8         & 35.92 & 16.00 & 0.10     & 0.032 &  & 3\\
  5-7-4-9019 & 06:55:15.30 & +10:38:59.0         & 33.87 & 15.92 & 0.34     & 0.015 &2  & 3\\
  5-7-4-7787 & 06:55:26.31 & +10:36:48.4         & 39.19 & 15.96 & 0.30     & 0.013 &  & 3\\
  5-7-4-7234 & 06:55:31.08 & +10:35:31.8         & 30.67 & 17.13 & 0.49     & 0.015 &  & 3\\
  5-7-4-3601 & 06:56:04.22 & +10:34:35.3         & 28.62 & 17.01 & 0.48     & 0.021 &  & 3\\
  5-7-3-1149 & 06:56:33.59 & +10:45:41.4         & 33.87 & 19.02 & 0.17     & 0.036 &1  & 3\\
  3-8-1-21492& 17:54:29.79 & +01:37:52.2         & 29.77 & 21.38 & $-$1.23  & 0.057 &   &3 \\
  3-5-3-16324 & 17:59:58.53 & +01:50:28.5        & 38.63 & 19.46 & 0.49     & 0.020 &  & 3\\
  3-14-2-12275 & 18:00:02.54 & +00:34:08.4       & 34.76 & 19.53 & 0.38     & 0.034 &  & 3\\
  3-14-3-6993 & 18:01:06.36 & +00:54:18.1        & 36.91 & 21.16 & 0.17     & 0.097 &  & 3\\
  3-5-3-1384 & 18:01:11.05 & +01:46:36.4	 & 38.39 & 21.39  & 0.46    & 0.085 &  & 3\\
  3-11-2-3507 & 18:02:50.62 & +00:43:52.8 	 & 34.14 & 18.97 & 0.29     & 0.048 &  & 3\\
  3-11-3-14721 & 18:03:18.71 & +00:57:30.1 	 & 28.61 & 21.88 & 0.47     & 0.092 &  & 3\\
  3-11-3-14333 & 18:03:20.82 & +00:52:40.9 	 & 33.09 & 20.91 & 0.43     & 0.053 &  & 3\\
  3-11-3-14041 & 18:03:22.48 & +00:52:07.8 	 & 33.73 & 21.11 & 0.46     & 0.047 &  & 3\\
  3-11-3-176 & 18:04:45.09 & +00:53:26.4 	 & 37.53 & 21.72 & $-$0.12  & 0.148 &  & 3\\
  3-24-3-21598 & 18:16:48.03 & +06:16:11.4 	 & 39.49 & 21.54 & 0.50     & 0.121 &  & 3\\
  3-24-3-27252 & 18:17:41.11 & +06:39:30.2 	 & 33.88 & 21.95 & 0.48     & 0.144 &  & 3\\
  3-24-3-27068 & 18:17:41.80 & +06:39:55.2 	 & 33.88 & 19.17 & 0.47     & 0.019 &  & 3\\
  3-24-3-26839 & 18:17:42.61 & +06:39:47.5 	 & 37.50 & 20.11 & 0.39     & 0.031 &  & 3\\
  3-24-3-25379 & 18:17:47.71 & +06:42:21.7 	 & 37.62 & 19.63 & 0.50     & 0.022 &  & 3\\
  3-24-3-24553 & 18:17:50.35 & +06:40:36.7 	 & 38.22 & 19.37 & 0.36     & 0.027 &  & 3\\
  3-24-4-10602 & 18:17:52.98 & +06:31:33.2 	 & 37.04 & 21.55 & 0.47     & 0.084 &  & 3\\
  3-24-3-23361 & 18:17:54.38 & +06:41:38.7 	 & 38.22 & 20.92 & 0.33     & 0.074 &  & 3\\
  3-24-3-22275 & 18:17:58.05 & +06:42:55.7 	 & 38.97 & 19.86 & 0.48     & 0.026 &  & 3\\
  3-24-3-20681 & 18:18:03.13 & +06:39:50.2 	 & 25.16 & 19.78 & 0.30     & 0.024 &  & 3\\
  3-24-4-7636 & 18:18:12.79 & +06:26:23.7 	 & 29.67 & 22.38 & $-$0.31  & 0.107 &  & 3\\
  3-24-3-13813 & 18:18:26.67 & +06:44:01.8 	 & 37.62 & 21.45 & 0.36     & 0.067 &  & 3\\
  3-24-3-13658 & 18:18:27.31 & +06:47:06.2 	 & 36.70 & 19.33 & 0.46     & 0.024 &  & 3\\
  3-24-1-8755 & 18:18:32.46 & +06:23:19.9	 & 35.92 & 19.41 & 0.48     & 0.021 &  & 3\\
  3-24-3-11788 & 18:18:33.04 & +06:37:17.9 	 & 39.49 & 18.25 & 0.45     & 0.024 &  & 3\\
  3-24-3-11343 & 18:18:34.46 & +06:37:09.7 	 & 34.97 & 20.85 & 0.48     & 0.048 &  & 3\\
  3-24-3-11006 & 18:18:35.62 & +06:38:43.4 	 & 35.08 & 21.44 & 0.46     & 0.074 &  & 3\\
  3-17-2-5279 & 18:21:41.24 & +08:22:28.3 	 & 27.85 & 19.70 & 0.21     & 0.041 &  & 3\\
  3-27-1-24759 & 18:23:14.11 & +07:36:50.9 	 & 34.05 & 20.48 & 0.28     & 0.049 &  & 3\\
  3-27-3-9867 & 18:23:36.70 & +07:58:00.5 	 & 29.74 & 19.17 & 0.44     & 0.022 &  & 3\\
  3-27-3-9744 & 18:23:37.56 & +07:59:07.1 	 & 28.86 & 22.32 & 0.41     & 0.131 &  & 3\\
  3-25-3-11740 & 18:42:54.64 & +00:21:48.7 	 & 36.12 & 21.76 & 0.09     & 0.168 &  & 3\\
  2-11-4-41975 & 19:53:06.30 & +18:48:39.4 	 & 32.19 & 17.87 & 0.30     & 0.023 &2  & 3\\
  2-11-1-24882 & 19:53:35.94 & +18:38:08.7 	 & 25.46 & 18.86 & 0.14     & 0.034 &1  & 3\\
  2-3-4-4027 & 20:02:18.77 & +18:49:37.9 	 & 34.40 & 20.41 & 0.50     & 0.107 &  & 3\\
  4-16-3-16761 & 21:05:03.14 & +46:27:45.5 	 & 32.97 & 15.11 & $-$0.16  & 0.034 &2  & 3\\
  4-16-4-21946 & 21:05:12.19 & +46:20:38.3 	 & 31.82 & 14.98 & 0.22     & 0.022 &2  & 3\\
  4-6-4-11525 & 23:47:46.82 & +56:28:52.0 	 & 31.34 & 21.15 & 0.09     & 0.094 &  & 3\\
\hline
\multicolumn{9}{|l|}{$^{1}$Low signal-to-noise spectrum -- no obvious emission lines,$^{2}$A-type star spectrum}\\
\multicolumn{9}{|l|}{
 $^{3}$This paper}
\end{tabular}
\end{table*}

We searched for the 66 variables in Table~\ref{tab:bspv25} and \ref{tab:bspv40} in the Simbad database\footnote{The SIMBAD database is operated at CDS, Strasbourg, France} and found three matches with 5 arcsec of our target. One is HM Cnc, an AM CVn system \citep{ramsay02} and two are previously published sources which we found in the RATS data: a high amplitude pulsating subdwarf B star \citep{ramsay06} which is a rare hybrid object as it exhibits both gravity and pressure modes of pulsation \citep{baran10}; and a dwarf nova with high amplitude quasi-periodic oscillations in quiescence \citep{ramsay09}. Of the other sources; 22 have periods less than 25 min, of which 9 have a period of less than 10 min. The amplitudes of the short-period blue sources ranges from 0.012--0.284 mag with a mean of 0.066 mag and the $g'$ band brightness range from 14.9--22.4 mag with a mean of 19.6 mag.

\section{Spectra of variable sources}
\subsection{Observations}
We have a programme to obtain optical spectroscopic observations of variable sources found in our survey (cf. \S \ref{variable}). These data are low resolution spectra with which to determine the nature of sources. Most data were obtained with the 4.2-m William Herschel Telescope on La Palma using the dual-beam ISIS instrument and the R300B and R158R gratings giving a wavelength coverage from 3500--10000\AA. 

Typically the slit-width was 0.8 arcsec and based on the FWHM of arc lines we estimate the spectral resolution to be approximately 3 and 6\AA, for the blue and red spectra, respectively. Wavelength calibration was determined using Cu-Ar and Cu-Ne arc-lamps.

We have also used the ALFOSC instrument on the Nordic Optical Telescope for brighter targets and GMOS on the Gemini South telescope for the very faintest targets ($g'> 20$). Southern sources have been observed with EFOSC2 on the ESO 3.6m telescope at La Silla Observatory, Chile and the Grating Spectrograph on the 1.9-m Radcliffe telescope at the South African Astronomical Observatory.

We reduced these data using standard techniques and employ the packages \textsc{Molly}, \textsc{Pamela}\footnote{\textsc{Molly} and \textsc{Pamela} were written by T. Marsh and can be found at http://www.warwick.ac.uk/go/trmarsh} and \textsc{Starlink}\footnote{The Starlink Software Group homepage can be found at http://starlink.jach.hawaii.edu/starlink} packages \textsc{Figaro} and \textsc{Kappa}. We used optimal extraction \citep{horne86} and did not observe flux standards so normalised spectra by fitting a spline to the continuum. Arc lamp exposures were typically taken at the beginning, middle and end of each night and we calibrated each spectrum by interpolating between the arcs.

\subsection{Fits with model spectra}
\label{sec:fitting}
The spectra we obtain are primarily for identification purposes (e.g. is the variable star a main sequence star or a hydrogen-rich white dwarf). However, for those cases where we have reasonable signal to noise we are able to determine a number of properties of the star through spectral fitting.

In order to determine surface temperatures, surface gravity and metallicities we fit the spectra we have obtained with model atmospheres using the \textsc{Fits2b} fitting programme \citep{napiwotzki04}. For stars with DA white dwarf spectra we fit the Balmer lines using a grid of hot white dwarf models \citep{koester01}\footnote{A grid of hot white dwarf models was kindly supplied by Detlev Koester} with temperatures ranging from 6000--100000 K and log g from 5.5--9.5. For all other spectra we use ATLAS9 model atmospheres \citep{castelli04}.

%The classification which we have given individual sources are shown in Tab.~\ref{tab:bspv25} and \ref{tab:bspv40}. In Fig.~\ref{fig:spectra} we show an example of each of the normalised spectra of 2 different classes of short-period blue variables for which we have spectra of -- a pulsating white dwarf and a variable A-type star.

%A number of Balmer lines of the A-type star, RAT J210514.75+461541.9, are shown in Fig.~\ref{fig:spec-fits}. Overplotted is the best-fitting main sequence star model which has an effective temperature of $8760\pm80$ K and a surface gravity of $4.15\pm0.05$ dex. These are fairly typical for an A-type star

\begin{figure}
\includegraphics{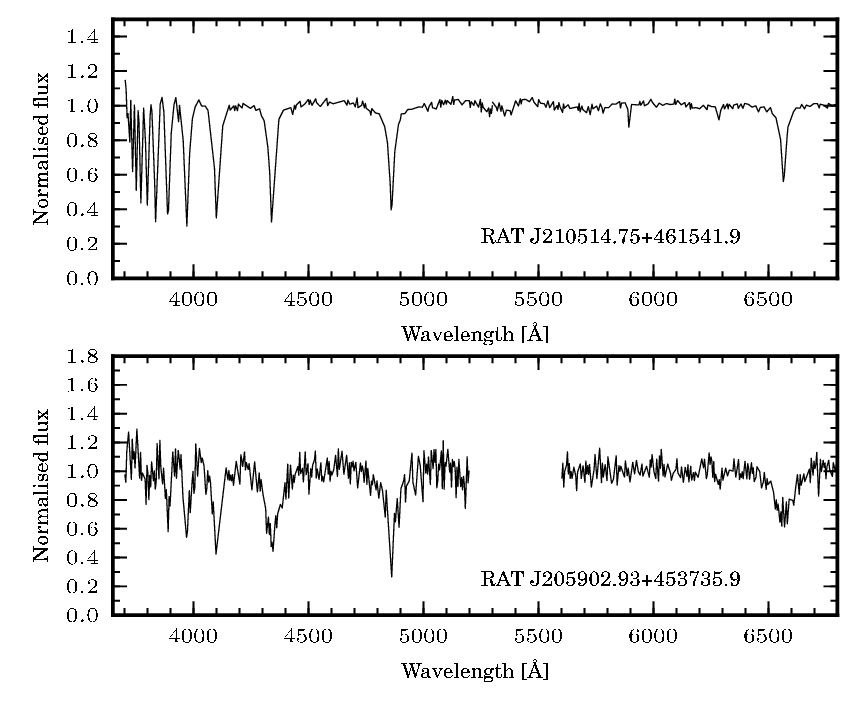}
\caption{Two optical spectra made using the ISIS spectrograph on the WHT. The upper spectrum is of a pulsating A star with a 22.4 min period in its light curve. The lower plot is of a pulsating DA white dwarf which has an optical modulation present in the light curve on a period of 15.3 min.}
\label{fig:spectra}
\end{figure}

\begin{figure}
\includegraphics{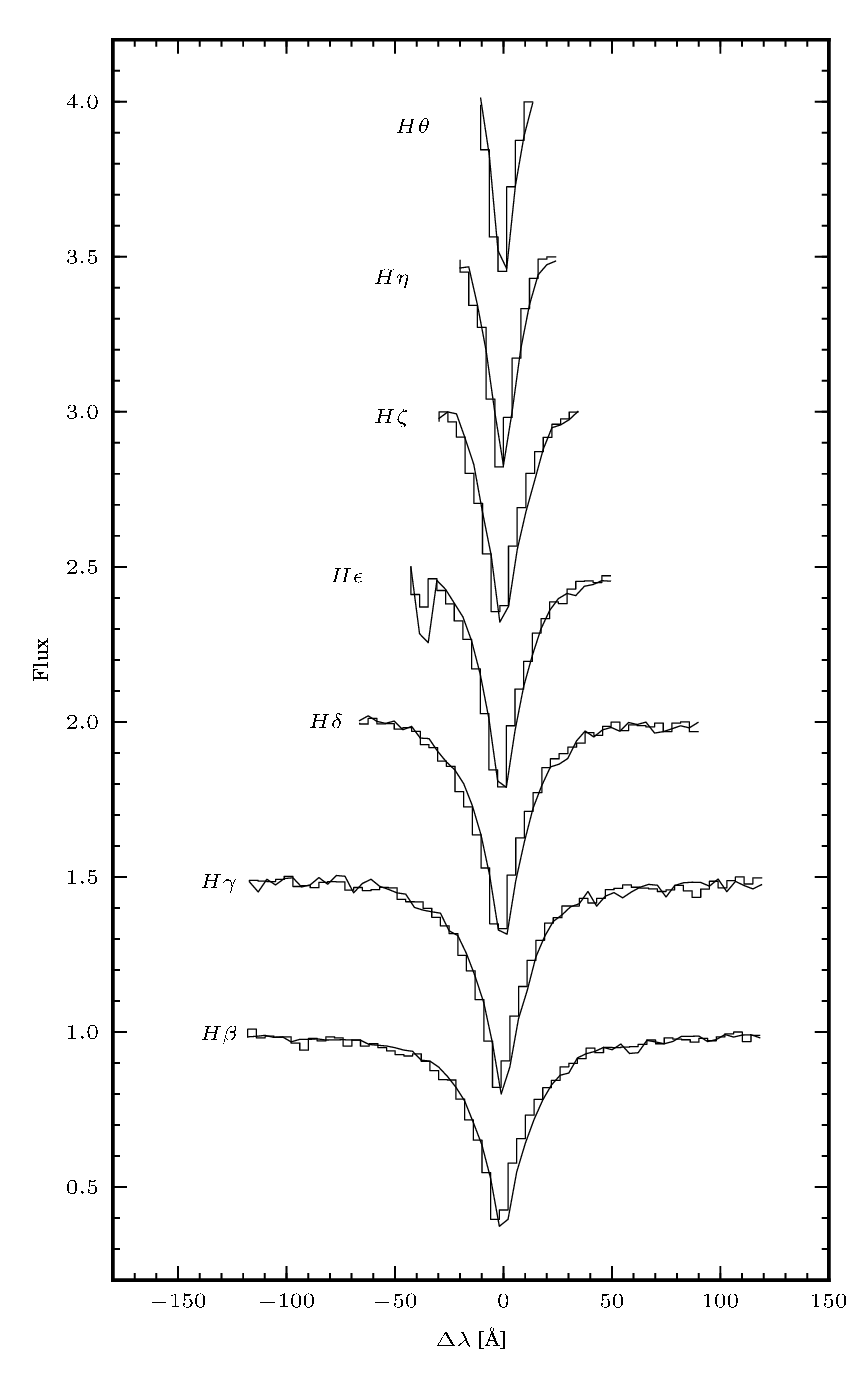}
\caption{The Balmer lines of RAT J210514.75+461541.9 over-plotted by the best-fitting model. The continuum of has been normalised to unity and each line after $H\beta$ has been shifted up by 0.5 in flux. The best-fitting effective temperature and surface gravity are $8760\pm80$ K and $4.15\pm0.05$ dex, respectively.}
\label{fig:spec-fits}
\end{figure}

%We have WHT spectra of 12 of the newly discovered short-period blue variables. One of these is a pulsating white dwarf which appears to have a hot degenerate companion (Barclay et al., in prep.). We have found 8 sources with the spectral type of a main sequence A star. One possibility is that they are long period examples of rapidly oscillating, chemically peculiar A (roAp) stars which have typical pulsation periods of 10 minutes \citep[e.g.][]{kochukhov09} although longer periods -- up to 21 minutes -- have been seen \citep{elkin05}. Alternatively they could be low amplitude delta Scuti stars with a pulsation period at the very short period end of the delta Scuti period distribution. We are planning medium-high resolution spectroscopic observation in order to distinguish between there two scenarios: roAp stars show evidence of heavy metals in their spectra, whereas the delta Scuti stars show abundances similar to the Sun.

We have WHT spectra of 12 of the newly discovered short-period blue variables. The classification which we have given individual sources as a result of this spectral fitting are shown in Table~\ref{tab:bspv25} and \ref{tab:bspv40}. One of these is a pulsating white dwarf which appears to be a composite of a WD pulsator of type DAV and a hotter companion (Barclay et al., in prep.), and 8 are sources with the spectral type of a main sequence A-type star. In Fig.~\ref{fig:spectra} we show the normalised spectrum of the pulsating white dwarf and of one of the variable A-type stars. The Balmer lines ($H\beta$--$H\theta$) of this A-type star, RAT J210514.75+461541.9, are shown in Fig.~\ref{fig:spec-fits}. Over-plotted is the best-fitting ATLAS9 model, which has an effective temperature of $8760\pm80$ K and a surface gravity of $4.15\pm0.05$ dex, where the errors were determined using a bootstrapping technique. We used a model with solar metallicity. These parameters are fairly typical for an A-type star. The nature of this, and the other 7 A-type variables, is unclear. One possibility is that they are long period examples of rapidly oscillating, chemically peculiar A (roAp) stars which have typical pulsation periods of 10 min \citep[e.g.][]{kochukhov09} although longer periods -- up to 21 min -- have been seen \citep{elkin05}. Alternatively they could be low amplitude delta Scuti stars with a pulsation period at the very short period end of the delta Scuti period distribution. We are planning medium-high resolution spectroscopic observation in order to distinguish between there two scenarios: roAp stars show evidence of a marked overabundances of heavy elements in their spectra, whereas the delta Scuti stars show abundances similar to the Sun.

\section{Space density of AM CVn systems}
\label{sec:spacedensity}
The combined sky coverage of the first 5 years of data is 31.3 square degrees. \citet{nelemans01,nelemans04} predict the space density of AM CVn systems with orbital periods less than 25 min as a function of Galactic latitude. This is done for a limit of $V<20$ and $V<22$ (for $B-V=0.2$, $V=g'$). For the distribution of our fields these models predict 2.7 and 8.2 AM CVn systems with $P_{orb} < 25$ min for $g'<20$ and $g'<22$, respectively.

In our survey we have identified 12 candidate AM CVn systems brighter than $g'=20$ and 33 brighter than $g'=22$ (cf. Table~\ref{tab:bspv25}). We have obtained optical spectra of 23 candidates, none of these spectra are consistent with the source being an AM CVn system. Removing these sources leaves us with ten candidates AM CVn systems for which we are yet to obtain spectra that are brighter than $g'=22$ of which six candidates are brighter than $g'=20$. One could speculate that the observed sample, once follow-up is completed, will indicate the need to revise the simulations, but for the time being we are left with an upper limit of ten, which is consistent with the model of \citet{nelemans01,nelemans04}. %We caution that these numbers are not highly constraining and we await the completion of the survey.

In contrast, the work of \citet{roelofs07} suggests that based on the number of systems discovered in the SDSS data,
the model of \citeauthor{nelemans01} over-predicts the
number of long period systems ($P_{orb}<25$ min) by a factor $\ga10$. If this were to be replicated at shorter periods we would expect to find $<1$ AM CVn systems in our observations so far. It remains to be seen whether there is also a deficit of systems
compared to the models at shorter periods, or whether there are
relatively many more younger, short period systems compared with older systems. This is
particularly important for future low-frequency
gravitational wave detectors such as \emph{LISA}.

\section{Conclusions}

Over the last decade, many surveys have set out to identify variable
objects in specific stellar clusters or nearby galaxies, or over
significant fractions of the sky, for instance in the search for
transiting exo-planets. However, very few of these surveys have been
able to identify sources with periodic variability on timescales as
short as a handful of minutes and in the range $15\le g\le22$. This brightness
range is important as the number of stars in this range is vastly
greater than at comparable brighter intervals, while still being
suitable for follow-up spectroscopy.

We have found that in a survey such as RATS it is essential that
potential sources of systematic variability are explored in detail.
Further, if resources permit, a number of pilot studies would result
in reducing the significance of these systematic effects. However,
once these issues have been dealt with, we have found that RATS is
very well suited to the discovery of variable stars with periods as
short as 5 min or show flares (or eclipses) as short as 5--10
min. Further, we can also identify stars with periods up to an hour
and stars which show variability over time-scales of several hours. Many of these sources are emerging as astrophysically
interesting. However, the sheer number of variable sources will
prevent us from following-up all the potentially interesting
sources. For this reason we intend to release to the community the
data products of the variable and non-variable sources presented in
this paper by the second half of 2011.

The inspiration for this project was to identify new ultra-compact
binaries. At this stage the implied space density of these objects is
consistent with the predictions of \citet{nelemans01,nelemans04}. However, we will not be in a position to make a definitive conclusion regarding
this question until we have fully analysed the final 16 square degrees
of data and obtained the essential follow-up spectroscopy of candidate
systems. It remains to be seen as to whether a study like this is the
optimum in identifying objects which appear to be intrinsically
rare.

\section*{Acknowledgements}

This paper is based on observations obtained using the INT and WHT
which are operated on the island of La Palma by the Isaac Newton Group
in the Spanish Observatorio del Roque de los Muchachos of the
Instituto de Astrof\'{i}sica de Canarias and also the MPG/ESO 2.2-m
telescope operated by the European Organisation for Astronomical
Research in the Southern Hemisphere, Chile (observing proposal number
075.D-0111). We thank the staff at both observatories for their
helpful advice and expertise. This research made use of tools provided by
Astrometry.net who we would like to thank. We used the \textsc{Pyfits}
libraries for handling FITS data in Python. \textsc{Pyfits} is a
product of the Space Telescope Science Institute, which is operated by
AURA for NASA. We made extensive use of the \textsc{Topcat}
\citep[][http://www.starlink.ac.uk/topcat/]{topcat} and
\textsc{Stilts} \citep[][http://www.starlink.ac.uk/stilts/]{stilts}
software packages. We thank Tom Marsh for the use of \textsc{Molly}
and \textsc{Pamela} software for the reduction of spectrum data.
Our thanks also go to Detlev Koester for providing us with DA white white dwarf model atmosphere. Balmer lines in these models were calculated with the modified
Stark broadening profiles of \citet{tremblay09},
kindly made available by the authors.

\bibliographystyle{mn2e}
\bibliography{rats-new}
\appendix \section{Fields Observed}

%The tables below are a summary of all the fields observed during the first 5 years of the RATS project.
\begin{table*}
\caption{Summary of the fields observed in the first 6 epochs of 
our survey. Field ID is in the
 format epoch-field\#, see table~\ref{tab:obs} for a summary of each
 epoch. Field centre refers to the centre of CCD4 in the INT WFC
 observations and close to the centre in the case of ESO WFC observations. The
 duration is the time-span for the series of 30 sec exposures. The seeing was determined
 from the FWHM of the images; the number of light curves is those stars 
 for which more than 60 data points were obtained.}
\label{tab:fields}
\begin{tabular}{lcccclr}
\hline
Field ID&Date& RA DEC& $l,b$ & Duration &Seeing &Number of
\\
        & & (J2000) & (J2000) & & ($''$) & lightcurves\\
\hline
INT1-2 & 2003-11-28& 22:57:08.9 +34:13:02& 97.5 $-$22.9 & 2h 33m& 0.8--1.3& 4061\\
INT1-6 & 2003-11-28& 02:08:10.7 +36:16:13& 139.8 $-$24.1 & 2h 23m& 1.0--1.5& 2786\\
INT1-11 & 2003-11-28& 04:11:31.1 +19:24:55& 174.6 $-$22.8 & 2h 04m& 1.0--1.5& 2762\\
INT1-21 & 2003-11-28& 07:29:20.0 +23:25:36& 195.4 +18.4 & 2h 31m& 1.0--1.5& 4097\\
INT1-1 & 2003-11-29& 23:10:43.9 +34:19:48& 100.3 $-$24.1 & 2h 47m& 1.2--2.0& 3464\\
INT1-8 & 2003-11-29& 02:02:15.6 +34:21:40& 139.2 $-$26.3 & 2h 22m& 1.3--2.6& 3321\\
INT1-16 & 2003-11-29& 04:55:55.5 +13:04:42& 186.9 $-$18.4 & 2h 02m& 0.9--1.3& 4783\\
INT1-806& 2003-11-29& 08:06:23.0 +15:27:31& 206.9 +23.4 & 2h 06m& 0.8--1.1& 4518\\
INT1-3 & 2003-11-30& 23:04:48.0 +34:26:19& 99.1 $-$23.4 & 2h 18m& 0.7--1.1& 4576\\
INT1-307 & 2003-11-30& 03:06:07.2 $-$00:31:14& 179.1 $-$48.1 & 1h 59m& 1.0--1.4& 1722\\
INT1-17 & 2003-11-30& 04:50:42.5 +18:11:59& 181.8 $-$16.4 & 2h 30m& 0.9--1.3& 4838\\
INT1-22 & 2003-11-30& 07:39 51.9 +23:50:03& 196.0 +20.8 &2h 41m& 0.9--1.3& 4644\\
\hline
INT2-1 & 2005-05-28& 13:57:08.6 +22:48:16&  20.3 +74.5 &  1h 53m& 1.5--3.0&  988\\
INT2-2 & 2005-05-28& 16:05:45.8 +25:51:45&  42.8 +46.8 &  1h 53m& 0.8--1.2&  2209\\
INT2-3 & 2005-05-28& 20:01:53.9 +18:47:41&  57.8 $-$6.2 & 1h 57m& 1.5--1.9& 32428\\
INT2-4 & 2005-05-29& 12:00:00.0 $-$00:00:00& 276.3 +60.2 & 0h 56m& 0.7--1.2&  1517\\
INT2-5 & 2005-05-29& 13:59:31.7 +22:10:56&  18.9 +73.8 &  1h 51m& 0.8--1.3&  1590\\
INT2-6 & 2005-05-29& 16:09:10.0 +24:00:13&  40.4 +45.6 & 1h 51m& 0.7--1.0&  2634\\
INT2-7 & 2005-05-29& 18:03:29.9 +29:56:25&  56.1 +22.8 & 2h 09m& 0.8--1.0&  4333\\
INT2-8 & 2005-05-30& 14:00:37.2 +22:45:59&  21.2 +73.7 & 1h 57m& 0.7--0.9&  2511\\
INT2-10& 2005-05-30& 17:59:12.6 +28:25:20&  54.2 +23.2 & 1h 51m& 0.6--0.8&  8465\\
INT2-11& 2005-05-30& 19:53:46.1 +18:46:42&  56.7 $-$4.6 & 1h 40m& 1.4--1.5& 90451\\
INT2-12& 2005-05-31& 13:58:40.1 +23:34:19&  23.5 +74.4 & 2h 26m& 0.6--0.8&  1983\\
INT2-13& 2005-05-31& 17:56:28.9 +29:09:39&  54.7 +24.0 & 2h 02m& 1.3--1.5&  8172\\
INT2-14& 2005-05-31& 19:55:30.5 +18:42:04&  56.9 $-$5.0 & 1h 57m& 1.3--1.5&  76748\\
\hline
ESO-1  & 2005-06-03 & 12:04:20  $-$24:50:31 & 289.6 +36.8 & 1h 46m & 1.0--1.3 & 1573\\
ESO-2  & 2005-06-03 & 14:03:32  $-$22:17:05 & 324.1 +37.6 & 1h 56m & 0.8--1.1 & 2872\\
ESO-5  & 2005-06-03 & 20:04:50  $-$24:04:46 & 17.9 $-$26.2 & 2h 32m & 0.8--1.2 & 8000\\
ESO-4  & 2005-06-03 & 16:23:36  $-$26:31:39 & 351.0 +16.0 & 3h 28m & 0.7--1.0 & 36424\\
ESO-7  & 2005-06-04 & 12:08:15  $-$22:56:35 & 290.1 +38.9 & 2h 37m & 1.4--2.6 & 1356\\
ESO-8  & 2005-06-04 & 13:53:46  $-$23:41:48 & 320.9 +37.0 & 2h 25m & 1.2--2.4 & 2567\\
ESO-10 & 2005-06-04 & 18:01:03  $-$26:54:22 & 3.5 $-$1.9   & 2h 16m & 0.7--1.2 & 142523\\
ESO-13 & 2005-06-05 & 12:02:40  $-$22:32:40 & 288.4 +38.9 & 2h 28m & 0.6--1.4 & 1696\\
ESO-14 & 2005-06-05 & 14:03:47  $-$24:55:05 & 323.1 +35.1 & 2h 27m & 0.7--1.1 & 4402\\
ESO-16 & 2005-06-05 & 18:07:54  $-$24:56:23 & 5.9 $-$2.3   & 2h 32m & 0.7--1.4 & 173696\\
ESO-17 & 2005-06-05 & 20:05:13  $-$22:32:36 & 19.6 $-$25.8  & 2h 35m & 0.8--1.4 & 8711\\
ESO-19 & 2005-06-06 & 12:01:38  $-$24:25:27 & 288.7 +37.1 & 2h 28m & 0.8--1.4 & 2556\\
ESO-21 & 2005-06-06 & 16:00:16  $-$25:33:01 & 347.9 +20.4 & 2h 27m & 0.6--0.9 & 9877\\
ESO-22 & 2005-06-06 & 18:36:26  $-$23:55:04 & 9.9 $-$7.6   & 2h 32m & 0.7--1.0 & 162417\\
ESO-6540 & 2005-06-07 & 18:06:01 $-$27:44:40 & 3.3 $-$3.3  & 2h 25m & 0.6--0.8 & 185173\\
ESO-25   & 2005-06-07 & 12:08:07 $-$25:14:10 & 290.7 +36.6 &1h 52m & 0.6--2.0 & 3512\\
ESO-30   & 2005-06-07 & 22:03:11 $-$24:43:56 & 26.8 $-$52.3 & 2h 00m & 0.6--0.9 & 2754\\
\hline
\end{tabular}
\end{table*}

\begin{table*}
\begin{tabular}{lcccclr}
\hline
Field ID&Date& RA DEC& $l,b$ & Duration &Seeing &Number of\\
        & & (J2000) & (J2000) & & ($''$) & lightcurves \\
\hline
INT3-2 & 2007-06-12& 17:59:00 +01:38:00& 28.4   +12.4&  1h 45m& 1.1--1.4&  34275\\  
INT3-3 & 2007-06-12& 19:42:00 +19:06:00& 55.6   $-$2.0 &  1h 19m& 1.0--1.3& 117341\\
INT3-4 & 2007-06-13& 18:04:00 +02:20:00& 29.6   +11.6&  2h 08m& 1.1-1.7&  13194\\
INT3-5 & 2007-06-13& 18:00:30 +01:35:00& 28.5   +12.0&  2h 00m & 0.9--1.2&  43524\\
INT3-6 & 2007-06-13& 19:41:00 +19:48:00& 56.1   $-$1.4&  1h 53m & 0.8--1.1& 135686\\
INT3-7 & 2007-06-14& 18:04:00 +01:31:00& 28.8   +11.2&  2h 05m & 0.9--1.5&  17583\\
INT3-8 & 2007-06-14& 17:55:00 +01:46:00& 28.0   +13.4&  2h 00m & 0.8--1.2&  34249\\
INT3-9 & 2007-06-14& 19:37:00 +19:47:00& 55.6   $-$0.6&  2h 00m & 0.8--1.0& 117956\\
INT3-10& 2007-06-15& 18:00:00 +02:13:00& 29.0   +12.4&  2h 16m & 1.3--2.6&  10358\\
INT3-11& 2007-06-15& 18:04:00 +00:46:00& 28.2   +10.9&  2h 20m & 1.0--1.4&  43084\\
INT3-12& 2007-06-15& 19:40:00 +22:50:00& 58.6    +0.3&  2h 00m & 0.8--1.1&  41726\\
INT3-13& 2007-06-16& 17:55:00 +02:22:00& 28.5   +13.6&  2h 06m & 1.3--1.8&  10323\\
INT3-14& 2007-06-16& 18:01:00 +00:42:00& 27.7   +11.5&  2h 00m & 1.2--1.5&  38365\\
INT3-15& 2007-06-16& 19:31:00 +19:03:00& 54.3    +0.2&  2h 04m & 1.1--1.5&  84418\\
INT3-16& 2007-06-17& 18:23:04 +05:53:25& 35.0    +9.0&  2h 17m & 1.3--2.0&   9553\\
INT3-17& 2007-06-17& 18:22:51 +08:25:34& 37.3   +10.3&  2h 10m & 0.9--1.2&  51788\\
INT3-18& 2007-06-17& 19:27:00 +22:40:00& 57.0    +2.8&  2h 09m & 0.9--1.0&  93095\\
INT3-20& 2007-06-18& 18:19:42 +05:52:01& 34.6    +9.7&  2h 04m & 1.9--3.0&   4622\\
INT3-21& 2007-06-18& 18:18:00 +07:35:41& 36.0   +10.9&  2h 00m & 1.4--2.6&  30470\\
INT3-22& 2007-06-18& 19:39:00 +20:24:00& 56.4   $-$0.7&  2h 00m & 1.1--1.4&  78067\\
INT3-24& 2007-06-19& 18:18:18 +06:31:15& 35.0   +10.3&  2h 00m & 1.1--1.5&  48472\\
INT3-25& 2007-06-19& 20:32:00 +25:11:00& 67.0   $-$8.5&  2h 00m & 1.2--1.7&  52974\\
INT3-26& 2007-06-20& 18:20:21 +08:11:47& 36.8   +10.6&  2h 00m & 1.0--1.6&  13554\\
INT3-27& 2007-06-20& 18:23:47 +07:51:23& 36.8    +9.7&  2h 00m & 0.9--1.2&  59027\\
INT3-28& 2007-06-20& 20:31:00 +27:26:00& 68.7   $-$7.0&  2h 00m & 0.8--1.1&  40099\\
\hline
INT4-1 & 2007-10-13&  21:04:25.9 +45:40:34&  87.2   $-$0.8& 2h 16m&   1.2--1.6& 52237\\ 
INT4-2 & 2007-10-13&  21:57:14.8 +54:01:00&  99.1   $-$0.6& 1h 56m&   1.3--1.7& 33917\\
INT4-3 & 2007-10-13&  01:27:01.0 +53:50:51& 128.2   $-$8.7& 2h 09m&   1.2--1.8&  9355\\
INT4-4 & 2007-10-13&  02:52:26.6 +50:42:29& 141.7   $-$7.7& 2h 55m&   1.1--2.0&  5707\\
INT4-5 & 2007-10-14&  21:01:15.0 +44:30:47&  85.9   $-$1.2& 2h 20m&   0.8--1.1& 51566\\
INT4-6 & 2007-10-14&  23:48:10.0 +56:26:08& 114.2   $-$5.4& 2h 29m&   0.9--1.1& 22154\\
INT4-7 & 2007-10-14&  01:42:15.0 +55:21:46& 130.2   $-$6.8& 2h 10m&   0.9--1.3& 16942\\
INT4-8 & 2007-10-14&  04:55:19.0 +34:42:02& 169.2   $-$5.5& 2h 19m&   0.9--1.2&  9030\\
INT4-9 & 2007-10-15&  20:59:15.0 +45:34:42&  86.5   $-$0.2& 2h 21m&   1.0--1.3& 57117\\
INT4-10& 2007-10-15&  23:48:04.0 +54:19:49& 113.7   $-$7.4& 2h 19m&   1.1--1.3& 17730\\
INT4-11& 2007-10-15&  01:41:42.0 +54:33:14& 130.2   $-$7.6& 2h 10m&   1.0--1.2& 15605\\
INT4-12& 2007-10-15&  05:03:31.0 +34:56:22& 170.0   $-$4.0& 2h 16m&   0.9--1.2& 12597\\
INT4-13& 2007-10-16&  21:08:06.0 +44:14:06&  86.5   $-$2.3& 2h 10m&   1.2--1.4& 37264\\
INT4-14& 2007-10-16&  23:48:04.0 +54:19:49& 113.7   $-$7.4& 2h 00m&   1.1--1.4& 11212\\
INT4-16& 2007-10-17&  21:05:39.0 +46:20:27&  87.8   $-$0.6& 2h 10m&   1.3--1.7& 56719\\
INT4-17& 2007-10-17&  01:26:19.0 +54:49:00& 128.0   $-$7.7& 1h 54m&   0.9--1.1& 12945\\
INT4-19& 2007-10-17&  05:04:51.0 +36:14:33& 169.1   $-$3.0& 2h 35m&   0.8--1.2& 14854\\
INT4-20& 2007-10-18&  21:01:49.0 +45:10:21&  86.5   $-$0.8& 2h 10m&   0.9--1.2& 33033\\
INT4-21& 2007-10-18&  23:57:34.0 +56:11:06& 115.4   $-$5.9& 2h 10m&   0.9--1.2& 23038\\
INT4-22& 2007-10-18&  01:42:22.0 +53:58:47& 130.5   $-$8.1& 2h 25m&   1.0--1.4& 13504\\
INT4-24& 2007-10-19&  21:07:34.0 +45:31:43&  87.4   $-$1.3& 2h 10m&   1.1--1.4& 52761\\
INT4-25& 2007-10-19&  23:52:37.0 +56:20:59& 114.8   $-$5.6& 2h 10m&   1.2--1.4& 16588\\
INT4-26& 2007-10-19&  02:54:43.0 +49:49:49& 142.4   $-$8.3& 2h 00m&   0.9--1.4& 10470\\
INT4-27& 2007-10-19&  05:07:38.0 +34:18:48& 171.0   $-$3.7& 2h 00m&   1.0--1.4& 10242\\
INT4-28& 2007-10-20&  22:09:27.0 +55:27:30& 101.3   $-$0.5& 2h 11m&   1.0--1.4& 43867\\
INT4-29& 2007-10-20&  00:02:06.0 +53:34:37& 115.6   $-$8.6& 2h 00m&   1.0--1.4& 14069\\
INT4-30& 2007-10-20&  02:49:23.0 +50:17:50& 141.5   $-$8.3& 2h 00m&   0.9--1.3& 12571\\
INT4-31& 2007-10-20&  05:03:57.0 +34:16:49& 170.6   $-$4.3& 2h 00m&   0.8--1.3& 12571\\
\hline
\end{tabular}
\end{table*}

\begin{table*}
\begin{tabular}{lcccclr}
\hline
Field ID&Date& RA DEC& $l,b$ & Duration &Seeing &Number of\\
        & & (J2000) & (J2000) & & ($''$) & lightcurves\\
\hline
INT5-2 & 2008-11-03& 01:49:58.6 +56:19:39& 131.0   $-$5.6&  2h 46m&     0.7--1.1&     9497\\
INT5-3 & 2008-11-03& 04:34:33.4 +39:55:22& 162.5   $-$5.2&  2h 31m&       0.7  &     7541\\
INT5-4 & 2008-11-06& 21:10:10.0 +49:31:56&  90.7   +1.0&  2h 10m&     0.8--0.9&    21967 \\
INT5-5 & 2008-11-06& 23:40:09.0 +57:01:33& 113.3   $-$4.5&  2h 49m&     0.8--1.0&    14379 \\
INT5-6 & 2008-11-06& 04:31:45.0 +40:27:12& 161.7   $-$5.3&  2h 11m&       0.7  &    10837 \\
INT5-7 & 2008-11-06& 06:55:57.5 +10:37:54& 203.9   +5.8&  2h 01m&     0.7--0.9&    11288 \\
INT5-8 & 2008-11-07& 23:24:50.2 +57:13:57& 111.4   $-$3.7&  2h 10m&     0.7--0.8&    15334 \\
INT5-9 & 2008-11-07& 03:05:05.7 +55:35:15& 141.1   $-$2.5&  2h 00m&       0.7  &    10605 \\
INT5-10& 2008-11-07& 06:04:33.8 +24:42:45& 185.8   +1.5&  2h 20m&       0.7  &    11340 \\
\hline
\end{tabular}
\end{table*}

\end{document}